\documentclass[%
,twocolumn%
,amssymb, amsmath,nobibnotes, aps, pre]{revtex4-1}
\usepackage{bm}%
\expandafter\ifx\csname package@font\endcsname\relax\else
 \expandafter\expandafter
 \expandafter\usepackage
 \expandafter\expandafter
 \expandafter{\csname package@font\endcsname}%
\fi

\usepackage{graphicx,subfigure}
\usepackage{amssymb,amsmath,wasysym}

\DeclareGraphicsExtensions{.pdf,.png}
\graphicspath{{./figs/}{./old_figs/}}
\usepackage[usenames,dvipsnames]{color}   

\usepackage[normalem]{ulem} 
\newcommand{\F}{F}

\def\be{\begin{equation}}
\def\ee{\end{equation}}
\def\bpm{\begin{pmatrix}}
\def\epm{\end{pmatrix}}

\renewcommand{\P}{\mathcal{P}}

\renewcommand{\L}{\mathcal{L}}

\DeclareMathOperator{\csch}{csch}

\begin{document}

\title{Active gel segment behaving as an active particle}
 


\author{P. Recho$^{1}$, T. Putelat$^{2,3}$ and L. Truskinovsky$^4$}
\affiliation{
$^1$LIPhy, CNRS--UMR 5588, Universit\'e Grenoble Alpes, F-38000 Grenoble, France\\
$^2$SAS, Rothamsted Research, Harpenden, AL5 2JQ, United Kingdom\\
$^3$DEM, Queen's School of Engineering, University of Bristol, 
Bristol, BS8 1TR, United Kingdom\\
$^4$PMMH, CNRS--UMR 7636, ESPCI PSL, F-75005 Paris, France
}
\email{\mbox{pierre.recho@univ-grenoble-alpes.fr, t.putelat@bristol.ac.uk,}  lev.truskinovsky@espci.fr}


\date{\today}%

\begin{abstract} \small 
 We reduce a one-dimensional  model of an  active segment (AS),  which is used, for instance,  in the description of contraction driven  cell motility  on tracks,  to  a zero-dimensional model of an active particle (AP) characterized  by two internal degrees of freedom: position and polarity.  Both models give rise to  hysteretic  force-velocity relations showing that an active agent can support two opposite polarities under the same external force and that it can maintain the same polarity while being dragged by  external forces with   opposite orientations. This double bi-stability results in a rich dynamic repertoire which we illustrate by studying  static, stalled, motile  and periodically re-polarizing  regimes  displayed by an active agent confined in a  visco-elastic environment. We show that the AS and AP models can  be calibrated to generate quantitatively similar dynamic responses.
\end{abstract}

\maketitle

\section{Introduction}

Most mamallian cells have a remarkable ability to self-propel even if confronted  by an opposing mechanical force~\cite{aranson2016physical}.  The implied  \emph{macroscopic} motion is generated \emph{microscopically}  inside the cellular cytoskeleton,   an actively crosslinked biopolymer meshwork  that can, for instance,  spontaenously and inhomogeneously contract in response  to various external and internal  stimuli \cite{Alberts2002}.  In particular, cells are  known to adjust their mode of self-propulsion  by sensing  the gradients of   chemokines, detecting the  density of   ligands and  probing   the stiffness of the environment \cite{holmes2017mathematical, ParkE5750}.  The  integration of all these cues \cite{Huang2004} allows cells to continuously reorganize their cytoskeleton and in this way to actively control their motility mechanism~\cite{lara2013directed, ziebert2013effects}. 

The effect of  mechanical stimuli  on the  dynamics of the cytoskeleton is raising an increasing interest \cite{iskratsch2014appreciating}. Some cells, like immune or cancer cells,  are typically  exposed to spatially inhomogeneous   rheological environments  which may generate  time dependent elastic and viscous resistence when they migrate in an organism.  Some cellular responses  can also be  directly  linked  to the  action of  external forces as in the case of the various outcomes of cell collision tests ~\cite{desai2013contact, scarpa2013novel}, which can be explained mechanically \cite{recho2019force} without   involving biochemical   pathways \cite{merchant2017rho,kulawiak2016modeling}. Understanding the response of the cytoskeletal reorganization to mechanical loading \cite{RecTru_pre13,putelat2018mechanical} may also  guide the    design  of micro-scale bio-inspired robots which would then assist various healing functions.

Many important advances have been made  in the  modeling of the migration of individual cells which involves not only cytoskeletal contraction but also other   complex phenomena, in particular, active  polymerization and active adhesion  \cite{mogilner2009mathematics}.  All these  mechanisms have been successfully captured by  the   continuum   liquid crystal theory with incorporated  tensorial chemo-mechanical coupling  \cite{JulKruProJoa_pr07,rubinstein2009actin,shao2010computational,ziebert2011model, giomi2014spontaneous,tjhung2015minimal}. However, while being comprehensive, the resulting models necessitate large scale numerical simulations. This  \textcolor{black}{makes their integration as  building blocks of a kinetic theory of tissues  \cite{Vicsek1995novel,peshkov2014boltzmann} computantionally costly and  simpler models are needed to study the  collective behavior of cells \cite{marchetti2013hydrodynamics}}.  Capturing  the  mechanical interaction of a  cell with its environment  at such \emph{reduced}  level   is crucial for the adequate reproduction of the   emerging  active phases~\cite{camley2017physical, hakim2017collective}.

To this end, we reduce in this paper  a one-dimensional  model of an  active segment (AS)~\cite{RecPutTru_prl13},  which is used, for instance,  in the description of contraction driven  cell motility  on tracks,  to  a zero-dimensional model of an active particle (AP) characterized  by two internal degrees of freedom: position and polarity.  By focusing on contraction, we are motivated by the experimental observations that a crucial building block of cell re-polarization, which plays an important role in  both cell collisions and cell oscillations,   is myosin contractility \cite{RecPutTru_prl13, barnhart2015balance}. In this explarotory study we limit our consideration to  one dimension having in mind that  such  setting is close to  the  classical well-calibrated experimental assays \cite{maiuri2012first} while also  carrying some physiological significance:  a typical situation of three-dimensional in-vivo motility  is when cells travel along the fibers of the extra-cellular matrix.  

 In contrast to some well known representations of size-less active agents~\cite{romanczuk2012active},  the obtained AP   model accounts for the temporal dynamics of the degree of    cell polarization. We present a systematic study of how such internal variable is affected by the time dependent external  forces. In particular, we show that  both AS and AP models support two coexisting dynamic regimes: \emph{frictional}, when the active object is dragged by the force, and \emph{anti-frictional}, when it is dragging  the cargo.  The fact that the  system is able to  switch from one of these nonequilibrium steady states to the other through a hysteresis loop shows that re-polarization can emerge as a result of the direct self-organization of the cytoskeleton in response to a mechanical action without additional bio-chemical regulation.  
 
  In the case of  self-propulsion   in a viscous environment  we find a continuous  transition between  the static (no polarity) and the motile (two symmetry-related polarities) regimes at a critical activity threshold which becomes viscosity independent at sufficiently large viscosities.  In the case of elastic confinement, we identify  three dynamic regimes: static (no polarity), stalled  (two symmetry-related polarizaties) and oscillatory (periodically varying polarity). In a certain range of parameters, the theory predicts  a metastable coexistence between the stalled and the oscillatory regimes,  which opens the possibility of complex stop and go dynamics in the presence of noise. 
  
We show that the AS and AP models can be  calibrated to generate not only  qualitatively  but also quantitatively similar dynamic responses. This is rather remarkable in view of the fact that the AS model is described by a free boundary problem formulated for nonlinear partial differential equations of Keller-Segel type while the AP model ultimately reduces to a single ordinary differential equation.

The paper is organized as follows.  In Sec.~\ref{sec:act_segment} we present the  AS model in the presence of a general    external force field. In Sec.~\ref{sec:act_paticle}, we formally reduce this model to a set of two ordinary differential equations describing a size-less active particle and specify the calibration procedure for the reduced model. In Sec.~\ref{sec:FV}, we compare the velocity-force relation obtained in AS and AP models and show that they can be made  quantitatively similar. Then in Sec.~\ref{sec:viscous}  we study   the dynamics of an AP subjected to a viscous force. The case of an AP  attached to fixed wall through a linearly elastic spring is studied  in Sec.~\ref{sec:elastic}.   Sec.~\ref{sec:conclusion} summarizes our results.

\section{The active segment (AS)}\label{sec:act_segment}

In this section we review the model of an active gel segment performing a contraction driven crawling on a rigid surface \cite{RecPutTru_prl13, RecPutTru_jmps15, aranson2016physical}.  Our focus is on the unexplored role of the distributed external forces in this context.

\subsection{\textcolor{black}{The active gel model}}

A cell crawling on a straight frictional substrate is represented as a viscous contractile gel of fixed length $L$. The mechanisms fixing the cell length (see \cite{putelat2018mechanical}) are not described here; this simplification is made to make the analysis  more  transparent.  

The time dependent free boundaries of the cell are $x_f(t)$ for the front and $x_r(t)=x_f(t)-L$ for the rear. The motion of the geometric center of the cell is described  by the function  $S(t)=(x_r(t)+x_f(t))/2$. For convenience, the actual position $x\in [x_r(t),x_f(t)]$ of a point inside the cell will be replaced in what follows by the traveling wave coordinate $y(x,t)=x-S(t)\in [-L/2,L/2]$. 

Momentum balance for the cytoskeleton meshwork with a frictional substrate requires that 
\begin{equation}\label{e:f_bal_dim}
\partial_y\sigma +f_e=\xi w,
\end{equation}
where $\sigma(y,t)$ is the axial stress field, $w(y,t)$ is the internal flow of the gel in the laboratory frame of reference, $\xi$ is a friction coefficient and  $f_e(y,t)$ is an external force field. The   resultant applied traction is therefore 
$$F_e(t)=\int_{-L/2}^{L/2}f_e(y,t)dy.$$ 

The constitutive behavior of the visco-contractile gel reads 
\begin{equation}\label{e:constitutive_dim}
\sigma=\eta \partial_y w+\chi c,
\end{equation}
where $\eta$ is the gel viscosity, $\chi$ is the contractility and $c(y,t)$ is the concentration of motors generating the active stress.

Since the segment  boundaries are impermeable to the gel, they are propelled at the common but unknown velocity,
\begin{equation}\label{e:impenetrability}
V(t)=\dot{S}(t)=\dot{x}_f(t)=\dot{x}_r(t)=w(\pm L/2,t),
\end{equation}
where the superimposed dot denotes the time derivative. The reaction stress at the two boundaries
$
\sigma(\pm L/2,t)=\sigma^b(t)
$
is a kinematic variable to be determined using the fixed length constraint.

The molecular motors are advected with the flow and also undergo a diffusive flux $J $ such that the motor conservation law reads
\begin{equation}\label{eq:motor_conserv}
\partial_tc+\partial_y[c (w-V)-J]=0.
\end{equation}
In accordance with  Fick's law we postulate that $J=D\partial_yc$ where $D$ is an effective \cite{RecPutTru_jmps15,aranson2016physical} diffusion coefficient. \textcolor{black}{The additional drift velocity $V$ in \eqref{eq:motor_conserv} is due to the fact that the time derivative is taken at fixed value of $y$ and $\partial_t|_{x \text{ fixed}}=\partial_t|_{y \text{ fixed}}-V\partial_y$.} Assuming the initial condition  $c(y,0)=c^0(y)$ and adopting no flux boundary conditions  
$
\partial_yc(\pm L/2,t)=0, 
$
we obtain that the total  amount of motors remains fixed
$$M=\int_{-L/2}^{L/2}c^0(y)dy=\int_{-L/2}^{L/2}c(y,t)dy.$$ 

\subsection{\textcolor{black}{Thermodynamics}}

A detailed  study of the thermodynamics of the AS model can be found  in \cite{RecJoaTru_prl14, aranson2016physical}.  Here, we present a simplified analysis in order to compare it with the case of an active particle. Assuming that temperature remains constant the  global dissipation in the system  $R$ reads 
 \begin{equation}\label{R2} 
R=P-\dot{\mathsf{E}}\geq 0,
 \end{equation}
where  $ \mathsf{E}$  is  the   energy   of the system and $P$ is the power of external forces. In view of \eqref{e:f_bal_dim} we can write
$$P= -\int_{-L/2}^{L/2}(\xi w-f_e)w dy.$$  
To compute  $\dot{\mathsf{E}}$, we need to take into account  the  chemical reaction supporting the  activity of the motors.  If  $\zeta(y,t)$ is the reaction progress variable  we write  (see  \cite{RecJoaTru_prl14} for details):
\begin{equation}\label{e:free_energy}
\dot{\mathsf{E}}=-\int_{-L/2}^{L/2}\left[ J\partial_y\mu+A\dot{\zeta}\right]dy ,
\end{equation}
where $A$ is the affinity of the reaction which is a prescribed constant measuring the degree of the non-equilibrium \cite{prost2015active} and  $\mu(c)$ is the chemical potential of the motors.  Under these assumptions, we obtain 
the explicit expression for the  dissipation 
$$R=\int_{-L/2}^{L/2}\left[ \sigma\partial_yw+J\partial_y\mu+A\dot{\zeta}\right]dy. $$

We now make the standard Onsager relations \cite{de2013non}  $J=l_{33}\partial_y\mu$ and introduce a coupling between mechanics and chemistry   in the form
$\sigma=\eta\partial_y w+l_{12}A $, $ \dot{\zeta}=-l_{12}\partial_y w+l_{22}A $.
A simple way to express the fact that the molecular motors play the role of a catalyst for the reaction is to assume that the related kinetic coefficients are proportional to the concentration $l_{12}=a c$ and  $l_{22}=b c$, where $a$ and $b$ are  constants. With this assumption, we recover the constitutive relation \eqref{e:constitutive_dim} with   $\chi =aA$. A second consequence is the mechanical feedback to  kinetics 
$\dot \zeta  = c (b A-a\partial_yw)$. Finally to recover the Fickian diffusion postulated to close the conservation law \eqref{eq:motor_conserv}, we need to assume a linear dependence of the chemical potential in the concentration field $\mu=kc$ and set $D=kl_{33}$. As a result, we obtain
\begin{equation}\label{R1}
R= \int_{-L/2}^{L/2} \left[bA^2c+\eta (\partial_yw)^2+kD(\partial_y c)^2\right] dy\geq 0.
\end{equation}
 The first term in \eqref{R1} describes dissipation due to chemical reaction, the second term is the viscous dissipation and the last term is the contribution due to diffusion.

\subsection{\textcolor{black}{Non-dimensionalization}}

We non-dimensionalize distances by the hydrodynamic length $\bar{l}=\sqrt{\eta/\xi}$, times by $\bar{t}=\bar{l}^2/D$, concentrations by $\bar{c}=M/L$ and stresses by $\bar{\sigma}=\xi D$ (and hence forces by $\bar{f}=\bar{\sigma}/\bar{l}$ and velocities by $\bar{w}=\bar{l}/\bar{t}$).  The ensuing problem depends on the three non-dimensional parameters:
$$\mathcal{L}:=\frac{L}{\bar{l}}\text{, }\F:=\frac{F_e}{\bar{\sigma}}\text{ and }\mathcal{P}:=\frac{M\chi}{\bar{l}\bar{\sigma}}.$$ 
 These parameters represent, respectively, the length of the segment  in the units of  the hydrodynamic length (i.e. the length over which a perturbation in the flow propagates before it is damped), the normalized resultant traction force applied to the system, which is generically  a function of time, and the normalized contractility of the motors.
The problem also depends on the imposed non-dimensional force field $f(y,t)=f_e(y,t)/F_e(t)$  constrained by the condition $\int_{-\mathcal{L}/2}^{\mathcal{L}/2}f(y,t)dy=1$. 

Although we consider from now on only non-dimensional variables, for sake of clarity, we keep the same notations for the physical variables (i.e. time, space, stress, velocity and concentration).

\subsection{\textcolor{black}{Reduction to a single non-local equation}}

\begin{figure}[h!]
\centering
\includegraphics[scale=0.4]{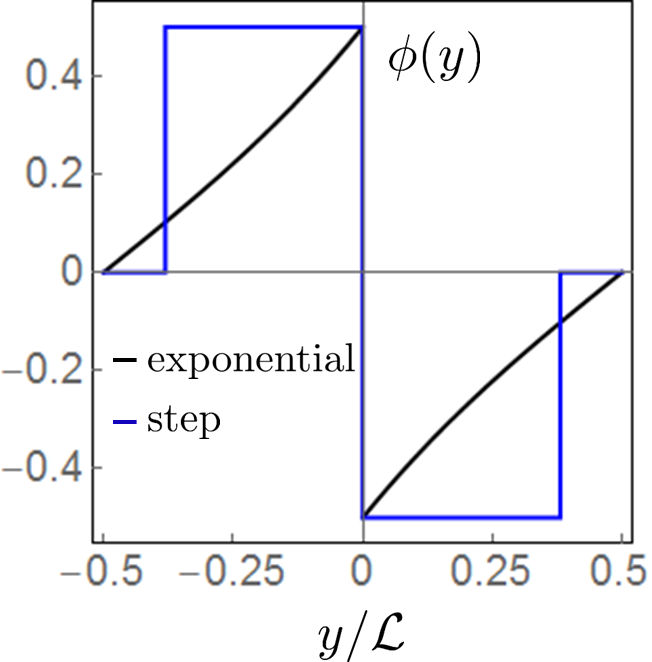}
\caption{(Color online) Two   interaction  kernels $\phi(y)$: black line is the exponential kernel of this paper and  blue line is  the kernel used in \cite{kruse2000actively}. Non-dimensional length $\mathcal{L}=2$. }
\label{f:kernel}
\end{figure}

Combining the force balance with the constitutive relation, we obtain   the linear equation for the stress
\begin{equation}\label{eq:force_bal}
-\partial_{yy}\sigma+\sigma=F\partial_yf+(\mathcal{P}/\mathcal{L}) c.
\end{equation} 
Solving for $\sigma$,  we  obtain a non-local relation  
\begin{equation}\label{eq:stress_non_loc}
\sigma=(\mathcal{P}/\mathcal{L})\tilde{\phi} \ast c+\F\tilde{\phi}\ast \partial_yf, 
\end{equation}
where we introduce  the  notation: 
$$\psi \ast h= \int_{-\mathcal{L}/2}^{\mathcal{L}/2}\psi(y-z)h(z,t)dz.$$
The  interaction kernel in \eqref{eq:stress_non_loc} is 
$$
\tilde{\phi}(z)=\frac{\cosh\left(z+\mathcal{L}/2\right)-2H(z)\sinh\left(z\right)\sinh\left(\mathcal{L}/2\right)}{2\sinh(\mathcal{L}/2)}, 
$$
where $H$ is  the Heaviside function. Differentiating the stress and using the force balance equation we  obtain the  expression for the velocity field
\begin{equation}\label{eq:velocity_non_loc}
w=(\mathcal{P}/\mathcal{L})\phi\ast c+ F(\phi\ast \partial_yf+  f),
\end{equation}
where $\phi(z)=\partial_z\tilde{\phi}(z)$.  In Fig.~\ref{f:kernel} we compare the kernel $\phi$  with the simplified kernel  introduced  in \cite{kruse2000actively} on purely topological and symmetry grounds. 

Equation \eqref{eq:velocity_non_loc} may be seen as as the fundamental description of the contraction-driven mechanics: the flow velocity $w$ at point $y$ is induced  first,  by  the presence in another point $z$ of an active force dipole, represented by a motor concentration-dependent active stress~\cite{RecPutTru_prl13,putelat2018mechanical}, and second,  by the  passive external force field.
\begin{figure}[h!]
\centering
\includegraphics[scale=0.6]{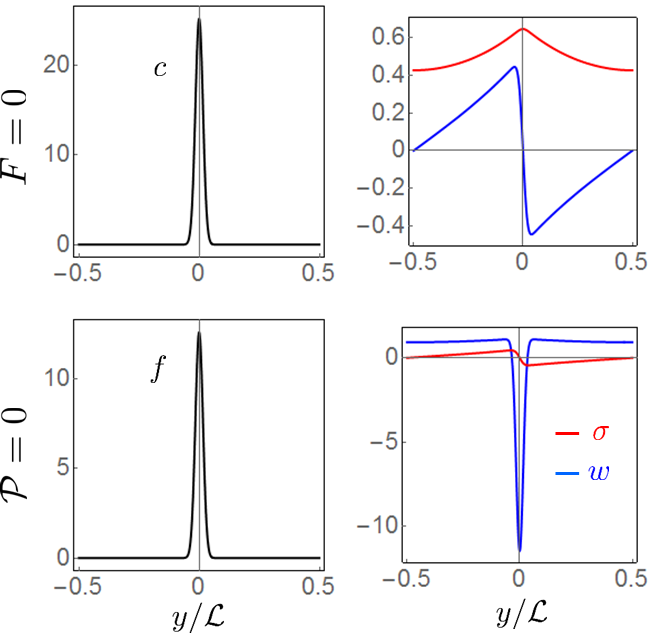}
\caption{\textcolor{black}{(Color online) Non local response of the stress field (see \eqref{eq:stress_non_loc}) and the velocity field (see \eqref{eq:velocity_non_loc}) to a   contractility distribution with no external force (first row, $F=0$ and $\mathcal{P}=1$) and to a   force distribution with no contractility (second row, $\mathcal{P}=0$ and $F=1$). Non-dimensional length $\mathcal{L}=2$.}}
\label{f:nonlocal}
\end{figure} 
\textcolor{black}{We illustrate in Fig.~\ref{f:nonlocal} the non-local response of the stress and velocity fields to a  space dependent motor/force loading. While a symmetric motor distribution gives rise to a symmetric stress field and an anti-symmetric velocity field, the  response to a symmetric force field is  the anti-symmetric stress distribution  and the symmetric velocity distribution.}

The  impenetrability condition \eqref{e:impenetrability} is then used  to express the segment  velocity   
\begin{equation}\label{eq:velocity_fronts}
V(t) = \frac{\P}{\L} \lbrace \phi \ast c\rbrace +\F\lbrace \phi \ast \partial_yf+f \rbrace,
\end{equation}
where $\left\lbrace h \right\rbrace =(h\vert_{-\mathcal{L}/2}+h\vert_{\mathcal{L}/2})/2$ denotes the average over the domain boundaries.

To write a single  equation for the concentration field we now combine \eqref{eq:motor_conserv}, \eqref{eq:velocity_non_loc} and \eqref{eq:velocity_fronts}  to obtain 
\begin{multline}\label{eq:model}
\partial_tc+ 
\partial_y \left[ c \left( (\P/\L)\phi \star c+\F(\phi \star \partial_yf + \delta \star f)\right) \right]  
= \partial_{yy}c,
\end{multline}
where $\delta$ is the Dirac distribution and we introduce the notation $\phi \star h=\phi \ast h-\lbrace \phi \ast h \rbrace$. The no-flux boundary condition then ensures that  
$ \langle c \rangle= 1$, where spatial averaging is defined by $$\langle h \rangle=\frac{1}{\mathcal{L}}\int_{-\mathcal{L}/2}^{\mathcal{L}/2}h(z,t)dz.$$
The field $c(y,t) $ must solve  \eqref{eq:model}.   In the special case when $F=0$ we obtain  \textcolor{black}{the classical Keller-Segel   model with a quadratic non-linearity \cite{Keller1971}
\begin{equation}\label{e:classical_KS}
\partial_tc+(\P/\L)\partial_y \left[ c \phi\star c \right]  
= \partial_{yy}c.
\end{equation}
}
After the equation \eqref{eq:model} is solved, the remaining unknowns $\sigma(y,t)$, $w(y,t)$ and $V(t)$ can be  reconstructed using equations \eqref{eq:stress_non_loc}, \eqref{eq:velocity_non_loc} and \eqref{eq:velocity_fronts}. Note that the velocity field decouples from the motor concentration field with the latter  emerging as the main driver of the overall dynamics.

\section{The active particle (AP)}\label{sec:act_paticle}

Suppose now  that the internal configuration of the motors $c(y,t)$ is not observable and that we only have  access to  some global polarity measure. We found  convenient  to choose it in the form 
$$C(t)=\left\lbrace \phi\ast c \right\rbrace/\mathcal{L},$$ 
which is a variable confined to the interval  $[-1/2,1/2]$ and which is non-zero if and only if $c$ is not symmetric (i.e. even). 

\subsection{\textcolor{black}{Model reduction}}
To obtain a closed description of the cell dynamics in terms of the `macro-variables' representing the polarity $C(t)$ and position $S(t)$, we need to project  the   infinite dimensional active segment (AS) model \eqref{eq:model} onto this two dimensional  space defining the active particle (AP) model. To this end, we first average \eqref{eq:velocity_non_loc} in two different ways. \textcolor{black}{Using  \eqref{eq:velocity_fronts} we directly obtain
$$V=\mathcal{P} C +  k_SF$$ 
where, 
\begin{equation}\label{e:k_Sgeneric}
k_S=\left\lbrace \phi\ast \partial_yf+f\right\rbrace.
\end{equation}}
By integrating \eqref{eq:velocity_non_loc} over space we also obtain 
$$\langle w \rangle= F/\mathcal{L}.$$ The new macroscopic variable $\langle w \rangle$ naturally enters the macroscopic analog of \eqref{eq:motor_conserv} which we write in the form $$\dot{C}+\langle w-V \rangle=-\Phi(C).$$ Here, the term $\langle w-V \rangle$ mimics the drift term in \eqref{eq:motor_conserv}. The term on the right-hand side is intended to play the role of diffusion degrading the  existing polarity and therefore the  function $\Phi$ is chosen to be increasing and vanishing at $C=0$. We thus write $\Phi(C)=\partial_C\bar E$, where the potential $\bar E$ is convex. In what follows we will be using the expression  
$$\bar E(C)=\frac{\alpha}{4} C^4+\frac{\mathcal{P}_c}{2}C^2.$$

If we  now eliminate $\langle w \rangle$ \textcolor{black}{and denote 
 $$k_C= k_S-1/\mathcal{L}, $$}  
we obtain the  system of ordinary differential equations:
\begin{equation}\label{eq:meta_model_231}
\begin{array}{c}
\dot S=\P C + k_S F \\
\dot C =-\partial_C E+k_C F,
\end{array}
\end{equation}
where we introduce  a  new  potential $E$ which now contains an active contribution: $\partial_CE=\Phi(C)-\mathcal{P} C$. In particular,  for the quartic choice of $\bar E$ made above, we obtain
$$E(C)=\frac{\alpha}{4} C^4-\frac{\mathcal{P}-\mathcal{P}_c}{2} C^2, $$  which is a Landau potential with the active term playing a destabilizing role for the symmetric state.  The presence of the active term $-\mathcal{P} C^2/2 $ ensures that for large $\mathcal{P}$ the potential develops two   wells corresponding to two symmetry related  polarized states. 

\textcolor{black}{Similar to the AS model, the  decoupling  of the  variable  $S$   from the dynamics of the variable $C$  renders the AP model \eqref{eq:meta_model_231}  non-potential: the position of the AP  depends on its polarity while  the reverse influence is absent.} 

\subsection{\textcolor{black}{Thermodynamics}}
If we multiply  \eqref{eq:meta_model_231}$_2$ by $\dot{C}$ we find,
\begin{equation*}
 k_C F\dot C-\dot E=\dot{C}^2\geq 0.
\end{equation*}
This  relation is reminiscent of \eqref{R2} in the AS model. The terms   $k_C F\dot C$ can be interpreted as the work done by the external force $F$ on the collective variable  $C$ representing the internal orientation of the particle.  The rate of change of the energy  associated with the variable $C$ is described by  the term   $\dot E$. Finally,    the positive definite term  $\dot C^2$ can be associated with dissipation $R$.  Note that   $\dot {E}$ splits into  the sum of a passive term $\dot{\bar{E}}$ representing diffusion and serving the same role as the term  $-\int_{-L/2}^{L/2} J\partial_y\mu dy $  in \eqref{e:free_energy}, while  the active term $-(\mathcal{P} C) \dot C$, representing the internally driven contraction, is the analog of the term $-\int_{-L/2}^{L/2}A\dot{\zeta}dy $.

\subsection{\textcolor{black}{Negative friction coefficient}}
\textcolor{black}{Given that the velocity of the particle is essentially enslaved to its polarity,  \eqref{eq:meta_model_231} is a direct analog of the Rayleigh-Helmholtz model  where the polarity variable is absent and the activity takes the form of a velocity dependent friction force \cite{romanczuk2012active}. Since the dissipation in such model can be negative,  the ``friction'' term will, in some parameter range, take the form of  an anti-friction, in particular, the   friction coefficient  becomes negative.}

\textcolor{black}{To illustrate this statement  it is sufficient to  compute the effective frictional viscosity  of the AP at zero velocity:
 $\mu_0(\P)=\partial_{V} F\vert_{\dot{S}=\dot{C}=0}$.
We obtain 
\begin{equation}\label{eq:side_forcing1}
\frac{1}{\mu_0}=k_S+\frac{\P}{\P_c-\P}k_C.
\end{equation} 
To highlight the effect,  consider   the simplest case when $\P_c\geq \P$ so that $E$ has a single well and \eqref{eq:meta_model_231}  has a single steady state ($\dot{S}=\dot{C}=0$) which is stable.  In the case of AS model, the results are  similar even though the analysis is much less explicit, see Appendix~\ref{sec:appendix_B}.}  

\textcolor{black}{The sign of $\mu_0$ in \eqref{eq:side_forcing1} depends on  the contractility $\P$ and the constants $k_S$ and $k_C$ which are functionals of the continuous force distribution.}

\textcolor{black}{One would expect that always $k_S\geq 0$ since in the absence of molecular motors ($\P=0$), a positive resultant force should be able to drag the layer in the forward direction.  When $f\geq 0$, we have indeed $k_S\geq 0$ but  negativity of $k_S$ can still result from a sign indefinite distribution of external loading, see Appendix~\ref{sec:appendix_C}.} 

\textcolor{black}{In contrast, even when $f\geq 0$, the coefficient  $k_C$ may be negative for some  force distributions. A negativity of  $k_C$ would mean that a positive resultant force favors negative polarity  which  triggers a competition between the active force $\P C$ and the passive force  $k_SF$ in determining  the AP velocity. On the contrary, a positive value of $k_C$ means that a positive value of the resultant force biases the polarity towards  a positive value and  that the active and passive forces conspire in selecting the velocity.}

\begin{figure}[h!]
\centering
\includegraphics[scale=0.75]{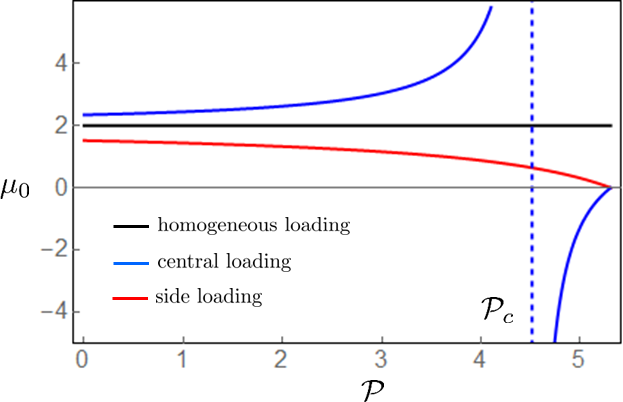}
\caption{(Color online) \textcolor{black}{Effective frictional viscosity $\mu_0$ in the  AP model as a function of the motor activity $\P$ in the three  loading configurations  (for the related AS model): homogeneous loading (black line),  loading localized in the middle of the segment (blue line),    loading on the segment sides (red line). Non-dimensional length $\mathcal{L}=2$.} }
\label{f:mu0}
\end{figure}

\textcolor{black}{We start with the simplest  situation   when the loading is homogeneous  
$f(y)=1/\L.$
In this case, we obtain $k_S=1/\L$ and  $k_C=0$. Thus, the coupling between the applied force and the polarity in \eqref{eq:meta_model_231} is absent  and the coefficient $\mu_0$ takes  its passive value (independent of $\P$) which is  $\mu_0=\L$, see the  black line in Fig.~\ref{f:mu0}. This is fully consistent with the behavior of the AS model as in the case of homogeneous loading \eqref{eq:model} is independent of the applied forces and reduces to \eqref{e:classical_KS}:  the homogeneous force only shifts the flow velocity $w$ by a constant  pulling the segment as if it was a passive object.}

\textcolor{black}{A more complex  case, which was also discussed in \cite{RecTru_pre13, recho2019force}, is  when external forces  are applied  at the  boundaries of the segment (for instance using cantilevers). Then 
\begin{equation}\label{eq:side_forcing}
f(y)=\beta\delta(y+\mathcal{L}/2)+(1-\beta)\delta(y-\mathcal{L}/2),
\end{equation}
where $0\leq\beta\leq 1$.  The configuration of the motors  is independent of the partition of the force between the two boundaries (factor $\beta$) because the length is fixed and the symmetric part of the loading on the boundary is absorbed into $\sigma^b$.  Therefore, independently of the value of $\beta$,  we obtain
\begin{equation}\label{e:kS}
k_S(\mathcal{L})=\frac{\coth(\mathcal{L}/2)}{2}.
\end{equation}
In this case  $k_C\geq 0$ and $\mu_0$ decreases with the motor activity reaching zero at $\P=\P_c$,  see the red line in Fig.~\ref{f:mu0}.}

\textcolor{black}{The situation changes radically in the case when  the loading is localized in the middle of the segment (imagine that a force is applied to the cell nucleus) 
$f(y)=\delta(y).$ Then $k_S=1/(2\sinh(\L/2))$ and thus $k_C\leq 0$. Since the coefficient  $k_C$ is negative, the friction coefficient $\mu_0$ increases with the motor activity $\P$ until it blows up and switches sign at the critical value  $\P=\L k_S\P_c$ from where it increases again to reach zero at $\P_c$, see the blue line in Fig.~\ref{f:mu0}. The fact that the frictional viscosity $\mu_0$ can reach zero and even be negative  is a  feature of many active systems \cite{haines2008effective, lopez2015turning}}.

\textcolor{black}{To summarize, the AP model carries a memory of the force distribution in the corresponding AS model and some particular force distributions may trigger the change of the sign of the effective friction coefficient.  In Section ~ \ref {sec:FV} we study the stationary force velocity distribution more systematically, showing, in particular,  how the friction coefficient  $\mu_0$ depends on the parameter $\P$.}

 In what follows, the force distribution will be always taken in the form \eqref{eq:side_forcing}.

\subsection{\textcolor{black}{Calibration}}

 To relate the AP and AS models quantitatively we need to find a relation between the functions $\phi(y)$  and $\Phi(C)$. To do so, it is sufficient to consider the case $F=0$. Under this condition, when the contractility parameter $\mathcal{P}$ in the AS model increases above a critical threshold $\mathcal{P}_c(\mathcal{L})$, the symmetric homogeneous solution of \eqref{eq:model}   $c\equiv 1$ becomes unstable and a polarized motile state emerges as a result of a pitchfork bifurcation (second order phase transition) leading to two symmetric configurations  with opposite polarities  \cite{RecJoaTru_prl14}.  The structure of the bifurcation is shown in Fig.~\ref{f:bifur_F0}  and the expression of $\mathcal{P}_c(\mathcal{L})$ is given in  Appendix~\ref{sec:appendix_A}. To reproduce the same bifurcation in the framework of the AP model (at $F=0$) , we  need to find the minima of the Landau potential 
$E(C)$. It has a single minimum at $C=0$ when $\mathcal{P}<\mathcal{P}_c$ and two symmetric minima at $C=\pm \sqrt{(\mathcal{P}-\mathcal{P}_c)/\alpha}$ when $\mathcal{P}>\mathcal{P}_c$, see Fig.~\ref{f:bifur_F0}. The coefficient $\alpha$ is fixed by matching the asymptotic behavior for the two models at $\mathcal{P}=\mathcal{P}_c$. From a normal form analysis of the AS model, we obtain $\alpha=\mathcal{P}_c^2\mathcal{L}^3\theta_2 (\mathcal{L})/2$;   the analytical expression for the function $\theta_2 (\mathcal{L})>0$ is given in Appendix \ref{sec:appendix_A}. 

\begin{figure}[h!]
\centering
\includegraphics[scale=0.45]{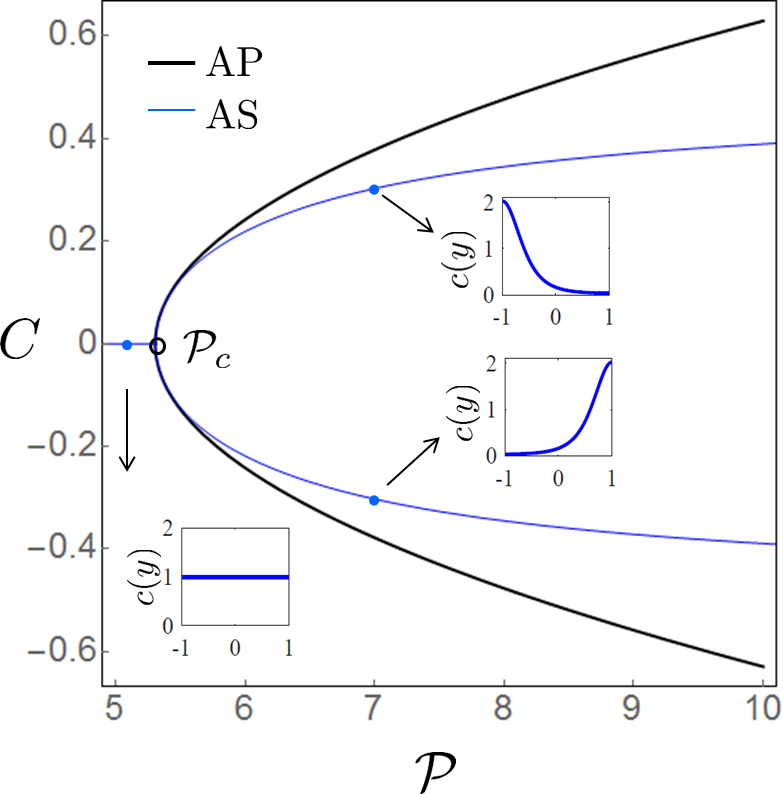}
\caption{ (Color online) Spontaneous polarization in  AS (thin blue line) and AP (thick black line) models when \textcolor{black}{contractility increases above the critical threshold $\mathcal{P}_c$}. We show in insets \textcolor{black}{ typical concentration profiles of molecular motors} along the bifurcated branches in the AS model. Non-dimensional length $\mathcal{L}=2$.}
\label{f:bifur_F0}
\end{figure}

\textcolor{black}{The last parameter that needs to be specified  to fully define the AP model is $k_S(\L)$ which encapsulates the external loading distribution. In this paper, we will focus on an external loading from the sides of the segment (see \eqref{eq:side_forcing}) leading to the expression \eqref{e:kS} for $k_S$. }  
  
The AP model is now fully defined and connected to the AS model by \eqref{eq:meta_model_231} with $\Phi(C)$, $\P_c(\L)$, $\alpha(\L)$ and $k_S(\L)$ given above. 

The dynamics of both the AP and AS thus depend on two scalar parameters $\P$ and $\L$ respectively characterizing the activity and size of the crawler and the external loading dynamic $\F(t)$.

\section{Velocity-Force relations}\label{sec:FV}

To test the efficiency of our calibration procedure, we now subject both systems, AS and AP, to a fixed external force and \textcolor{black}{show that the  steady state velocity-force \mbox{(V-F)} relations obtained in \cite{recho2019force} for the  AS model can be closely approximated if we use   directly the AP model} . 

\begin{figure}[h]
\centering
\includegraphics[scale=0.5]{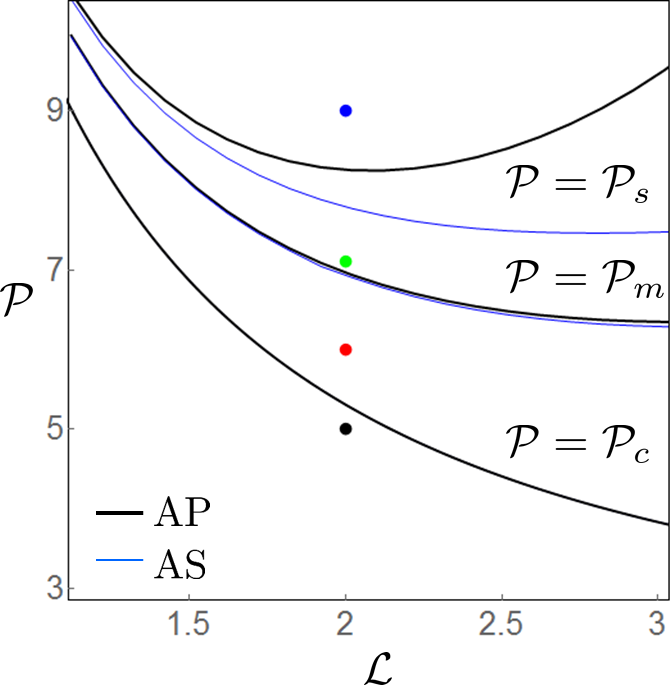}
\caption{(Color online) Dependence of three \textcolor{black}{contractility thresholds $\mathcal{P}_c$, $\mathcal{P}_m$  and $\mathcal{P}_s$ on the non-dimensional length $\mathcal{L}$ for the AP (thick black line) and AS (thin blue line).} The value of $\mathcal{P}_c$ is the same in both the AS and AP models by construction. The color dots are choices of parameters related to the V-F relations shown in Fig.~\ref{f:velocity_force2}.}
\label{f:velocity_force1}
\end{figure}

\begin{figure}[h]
\centering
\includegraphics[scale=0.5]{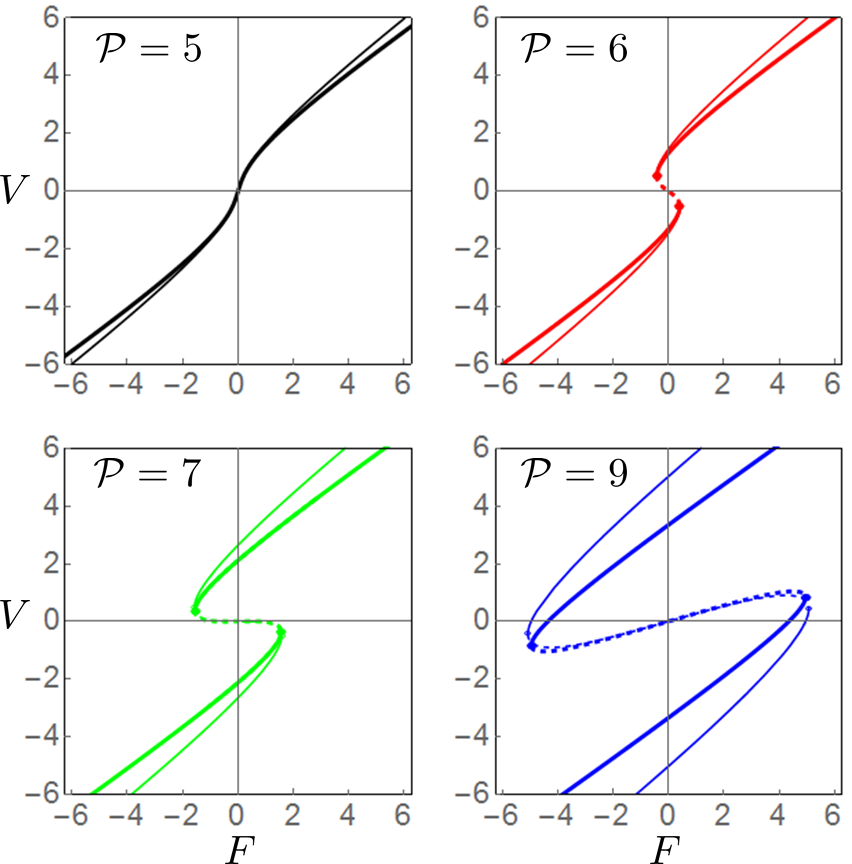}
\caption{\textcolor{black}{(Color online) Comparison of the V-F relations in the AS and AP models. Four typical V-F relations in the AS (thick lines) and the AP (thin lines) models.  The dashed parts of the V-F curves correspond to unstable regimes.  Parameters $\mathcal{L}=2$ and $\mathcal{P}=\textcolor{black}{5}$ (black-upper left corner),  $\mathcal{P}=6$ (red-upper right corner), $\mathcal{P}=7$ (green-lower left corner) and $\mathcal{P}=9$ (blue-lower right corner) are represented with the same color dots in Fig.~\ref{f:velocity_force1}.}}
\label{f:velocity_force2}
\end{figure}

In the case of the AS model, we solve  numerically equation \eqref{eq:model} with  $\partial_tc=0$. In the AP setting we find  the stationary value of polarity $C$  directly from the equation 
$\partial_CE= k_C  F$ and then obtain the V-F relation  substituting this value of $C$ into \eqref{eq:meta_model_231}.  As shown on Fig.~\ref{f:velocity_force2}, both models generate quantitatively similar V-F  relations. 

When \textcolor{black}{the contractility is sufficiently small,} $\mathcal{P}<\mathcal{P}_c$ \textcolor{black}{(black curves in Fig.~\ref{f:velocity_force2})}, the V-F relations in both models are single-valued and frictional, meaning that $V F >0$.   This is obvious  in the AP  case since  the potential $E(C)$ is convex  and the system has only one  stable ($\partial_{CC}E(C_0)>0$) stationary solution $C_0(F)$. The ensuing V-F  relation can be written explicitly:  $V =k_S F+\mathcal{P} C_0( F)$.

When \textcolor{black}{contractility becomes large enough,} $\mathcal{P}>\mathcal{P}_c$ \textcolor{black}{(red, green and blue curves in Fig.~\ref{f:velocity_force2})},  the V-F curves  develop a domain of bi-stability which spreads over a range $F\in[- F _t, F _t]$, where, in the AP model, $F_t=2(\mathcal{P}-\mathcal{P}_c)^{3/2}/(3k_C\sqrt{3\alpha})$.  Within this range, the stationary polarity can take three values: $C_0^* <C_0<C_0^{**} $  where $C_0^*<0<C_0^{**}$ correspond to metastable  solutions and $C_0$ is an unstable solution ($\partial_{CC}E(C_0)<0$). In this range, the \mbox{V-F} relations allow for  the coexistence of the two  metastable regimes with different signs of velocity:   $V^{*} =k_S F+\mathcal{P} C_0^{*}(F)$ and $V^{**} =k_S F+\mathcal{P} C_0^{**}(F)$. These two branches of the \mbox{V-F} relation are connected by the unstable branch $V_0=k_S F +\mathcal{P} C_0(F)$, which is located between the two turning points $F=\pm  F_{t}$. Inside  the coexistence interval $[- F_t, F_t]$, one of the two metastable solutions necessarily operates in an anti-frictional regime with $VF\leq 0$. Similar bi-directionality  is also characteristic of the V-F curves describing an  ensemble of molecular motors   interacting  either hydrodynamically  \cite{malgaretti2017bistability} or through a rigid backbone~\cite{julicher1995cooperative}. 

\begin{figure}[h]
\centering
\includegraphics[scale=0.7]{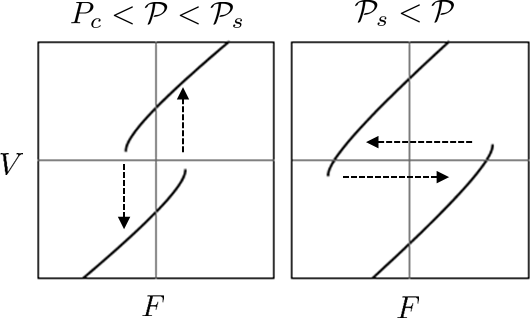}
\caption{\textcolor{black}{Schematic representation of the  single and  double hysteretic V-F relations.}}
\label{f:hysteresis}
\end{figure} 

The most interesting feature of the model is the existence of another threshold, $\mathcal{P}_m$ ($=k_S\mathcal{L}\mathcal{P}_c$ in  the AP case), beyond which  the V-F curves  start to display  muscle-like  stall force states. For $\mathcal{P}_m<\mathcal{P}<\mathcal{P}_s$ \textcolor{black}{(green curves in Fig.~\ref{f:velocity_force2})},  where $\mathcal{P}_s=2k_S\mathcal{P}_c/(3/\mathcal{L}-k_S)$ in the AP case, such states are unstable but for $\mathcal{P}>\mathcal{P}_s(\mathcal{L})$ \textcolor{black}{(blue curves in Fig.~\ref{f:velocity_force2})} they stabilize.
  The functions  $\mathcal{P}_{m,s}(\mathcal{L})$ for the AS model  are compared with those for the AP model in Fig.~\ref{f:velocity_force1} and the corresponding  V-F curves can be read off at Fig.~\ref{f:velocity_force2}. \textcolor{black}{As we illustrate in the schematic Fig.~\ref{f:hysteresis}, the V-F relations can display a standard hysteresis in force only (when $\mathcal{P}_c<\mathcal{P}<\mathcal{P}_s$) or be double hysteretic in both force and velocity (when $\mathcal{P}>\mathcal{P}_s$). In this case, not only two steady state velocities can be compatible with the same loading but also two force distributions  can be  compatible with the same velocity.}
  
\begin{figure}[h]
\centering
\includegraphics[scale=0.3]{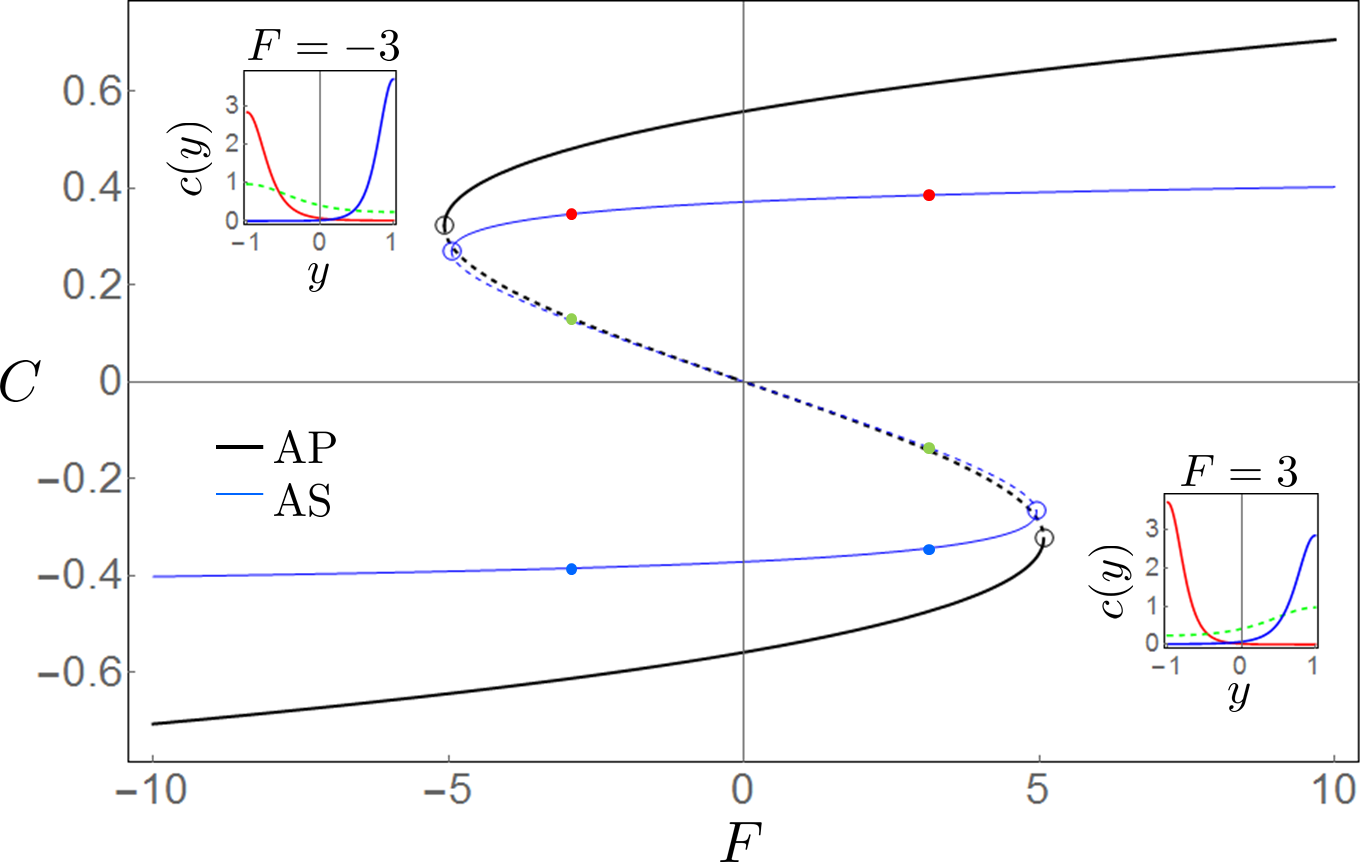}
\caption{(Color online) Comparison of the global polarity $C$ as a function of the force in AS (thin blue) and AP models (thick black).  The dashed parts of the C-F curve correspond to unstable regimes. We show in inset the motor concentration in the AS case at forces $F=-3$ and $F=3$. The red profiles have a positive polarity while the blue profiles have a negative polarity. A central symmetry transforms a red (resp. blue) profile at a positive force into a blue (resp. red) profile at a negative force. The green profiles are related to the unstable branch.  Non-dimensional length and contractility: $\mathcal{L}=2$ and $\mathcal{P}=9$.}
\label{f:polarity_force}
\end{figure} 
Note that in both the AS and AP models, the relation linking the global polarity measure $C$ to the velocity and the force is the same:
$$C=(V-k_SF)/\P.$$
We illustrate in Fig.~\ref{f:polarity_force} how $C$ varies as a function of $F$ and how the underlying concentrations of molecular motors change in the AS model. When the loading increases beyond the turning points located at $\pm F_t$, the global polarity changes sign as the local motor concentration  abruptly switches from one edge of the segment to the other. Similar hysteretic effects have also been found between the angular velocity and the applied torque in a Couette cell containing a polar active gel \cite{furthauer2012taylor}.
  
Given the good agreement between the AS and AP models in predicting the steady state regimes, in the rest of the paper we focus on the non-steady state dynamic behavior of the AP model in two paradigmatic cases when the AP is subjected to either a fluid-like viscous or a solid-like elastic environment. We also compare the non-steady behavior predicted by the AP and the AS models in the same conditions.

\section{Viscous environment}\label{sec:viscous}
\begin{figure}[h!]
\centering
\includegraphics[scale=0.55]{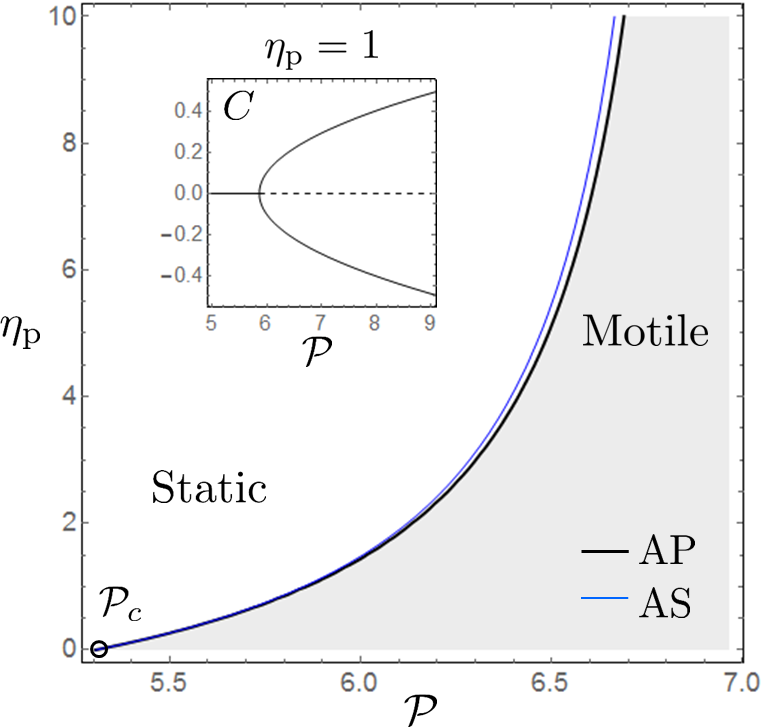}
\caption{(Color online) Phase diagram of an AP subjected to a viscous \textcolor{black}{drag force}. The inset shows that the bifurcation from a static to a motile case at a critical contractility remains of second order (supercritical). \textcolor{black}{Such transition can be also  obtained by  reducing the environment viscosity.} The phase boundary obtained for the AS is superimposed in thin blue. Non-dimensional length $\mathcal{L}=2$.}
\label{f:active_particle_viscosity}
\end{figure}

Assume that the external force is proportional to the particle velocity 
$$\F(t)=-\eta_{\text{p}}\dot{S}(t),$$
where $\eta_{\text{p}}$ is the (non-dimensional) viscosity of the   environment. 
System \eqref{eq:meta_model_231} then takes the form,
\begin{equation}\label{eq:viscosity}
(1+\eta_{\text{p}}k_S)\dot{S}=\P C\text{, } \dot{C}=-\partial_CE(C)-\eta_{\text{p}} k_C \dot{S}
\end{equation}
We can reduce \eqref{eq:viscosity} to a single non-linear ODE for  the particle polarity $\dot{C}=-\partial_CE_{\text{eff}}(C)$ where 
\begin{multline}\label{eq:viscosity_eff}
  E_{\text{eff}}(C)=\alpha \frac{C^4}{4}-\left( \mathcal{P}\left(1-\frac{\eta_{\text{p}} k_C}{1+\eta_{\text{p}}k_S} \right) -\mathcal{P}_c\right) \frac{C^2}{2}.
\end{multline}
The   viscosity of the environment  therefore  redresses the onset of motility   to the value
$$\P_c^{\text{eff}}=\frac{\P_c}{1-\frac{\eta_{\text{p}} k_C}{1+\eta_{\text{p}}k_S}}\geq \P_c.$$
The resulting motility initiation phase diagram is shown in  Fig.~\ref{f:active_particle_viscosity}. The effect is the same in the AS case with $\P_c^{\text{eff}}$ analytically given in Appendix.~\ref{sec:appendix_A} and presented for  comparison in Fig.~\ref{f:active_particle_viscosity} (thin blue line). Note that the bifurcation from a static to motile state as the activity of the motors increases remains supercritical,  see inset of Fig.~\ref{f:active_particle_viscosity}). 

Interestingly, the same type of transition (static to motile) is also initiated by reducing the environment viscosity.  However,  $\P_c^{\text{eff}}(\eta_{\text{p}}) $ has an asymptote ($=\P_m$ for both the AP and AS) when  $\eta_{\text{p}}\rightarrow \infty$. This is an indication that the transition threshold  $\P_c^{\text{eff}}$ depends  weakly  on the external viscosity when the latter  is sufficiently large   even though the velocity of the particle  remains sensitive to it.  The robustness of the threshold  suggests that, in this range of parameters  active crawlers can effectively adapt their   degree of polarization   to the external viscosity. 
\begin{figure}[h!]
\centering
\includegraphics[scale=0.5]{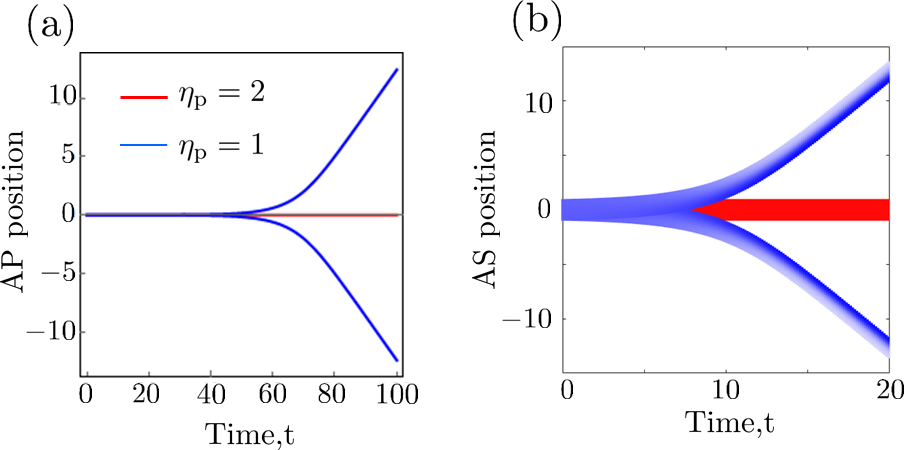}
\caption{ \textcolor{black}{(Color online) Initiation of motility} in a viscous environment. (a) AP position $S$ as a function of time with two external viscosities corresponding a static (red) and a motile (blue) case. One of the two symmetric trajectory is chosen according to a small bias in the initial polarity. (b) Same type of dynamics for the AS model. The intensity of the coloring is proportional to the level of motor concentration. Non-dimensional contractility and length $\P=6$ and $\mathcal{L}=2$.}
\label{f:active_particle_viscosity_dyna}
\end{figure}

In Fig.~\ref{f:active_particle_viscosity_dyna} we illustrate the nonsteady motility initiation phenomenon while comparing  the AP and AS dynamic models. One can see that depending on the value of the environmental  viscosity the same statically equilibrated  initial state can be stable or not: in more viscous environments active agents remain static while in less viscous environments they \textcolor{black}{spontaneously start}  to move in one of the two symmetric directions.

\section{Elastic confinement}\label{sec:elastic}
\begin{figure}[h!]
\centering
\includegraphics[scale=0.7]{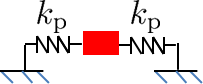}
\caption{Scheme of an elastically confined AP (in red).}
\label{f:elast_conf}
\end{figure}

We now consider the case where the environmental force $\F(t)$  is given by
$$\F(t)=-k_{\text{p}}S(t),$$
where  $k_{\text{p}}$ is the (non-dimensional) stiffness of the confining environment, see Fig.~\ref{f:elast_conf}. \textcolor{black}{While we have previously investigated this situation numerically in the case of  the  AS model \cite{recho2019force}, we now show  that the AP model allows one to  understand the stability properties of such systems analytically.}
Inside such a harmonic trap  the active agent  cannot move persistently but it can still have a rich dynamics. In the case of AP model,  the main   system \eqref{eq:meta_model_231}  takes the form,
\begin{equation}\label{eq:oscillations}
\dot{S}=\P C-k_{\text{p}}k_S S\text{, } \dot{C}=-\partial_CE(C)-k_{\text{p}} k_C S,
\end{equation}
which combines into the second order system for the polarity variable 
\begin{equation}\label{eq:oscillations1}
k_{\text{p}}\ddot{C}+(k_{\text{p}}\partial_{CC}E(C)+k_{S})\dot{C}+k_{S}\partial_{C}E(C)+k_{C}\P C=0.
\end{equation}

Inspection of equation \eqref{eq:oscillations1} shows that it reduces to a classical Van der Pol equation if $k_{S}=0$. The critical points of \eqref{eq:oscillations} are,
 $(S_0,C_0)=(0,0)$, corresponding to the force free static configuration 
and 
$$(S_s^{\pm},C_s^{\pm})=\pm\frac{\sqrt{k_S(\P-\P_c)-k_C\P}}{\sqrt{k_S\alpha}}\left(\frac{\P}{k_{\text{p}}k_S},1 \right), $$
describing  two symmetrically stalled configurations with  the spring  either under tension or compression. The  linear stability of such states is determined by solving the characteristic equation 
$$
\det\left(
\begin{array}{cc}
 -k_S k_p-\omega  & \P  \\
 -k_C k_p & -3 \alpha  C^2+\P -\P_c-\omega  \\
\end{array}
\right)=0
$$ 
for $C=C_0$ and $C=C_s^{\pm}$ and finding conditions when the  real part of $\omega$ becomes positive.  

From such  analysis, we find that the loss of linear stability of the trivial \textcolor{black}{static configuration $(S_0,C_0)$  can be of two  types depending on the rigidity of the external environment}:
\begin{itemize}
\item If $k_{\text{p}}\leq k_{\text{p}}^*=k_C\L\P_c/k_S$, the configuration $(S_0,C_0)$ stops being  linearly stable as a result of a Hopf bifurcation taking place at  $\P= k_{\text{p}}(\P_m-\P_c)/k_{\text{p}}^*+\P_c$.
\item If $k_{\text{p}}\geq k_{\text{p}}^*$, the configuration $(S_0,C_0)$ stops being  linearly stable through a supercritical pitchfork bifurcation taking place at  $\P= \P_m$.
\end{itemize} 

\begin{figure}[h!]
\centering
\includegraphics[scale=0.45]{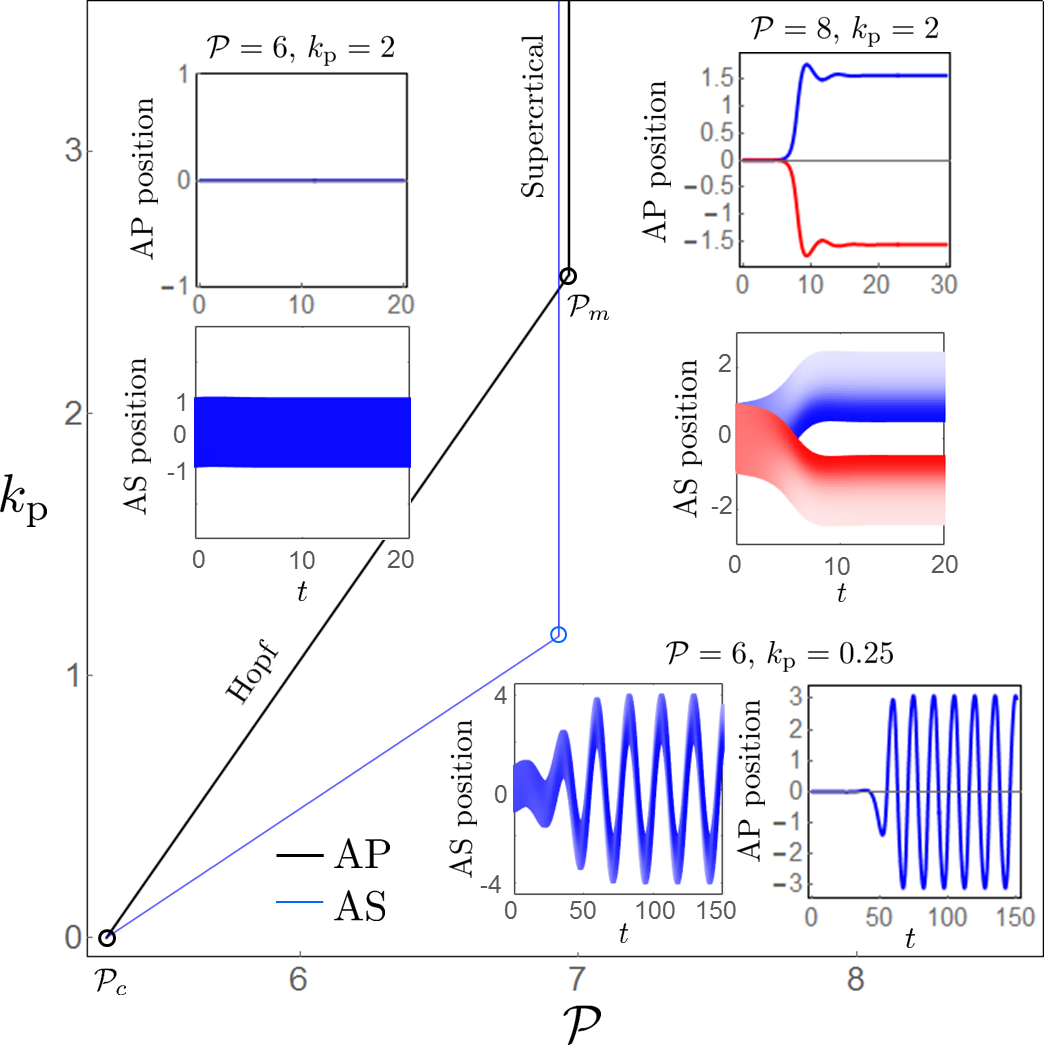}
\caption{ (Color online) Linear instability thresholds of the static solution in the AP and AS case. We show in insets some typical static, stalled and oscillatory dynamics for the AP and AS case. The intensity of the coloring is proportional to the motor concentration for the AS case. Non-dimensional length $\mathcal{L}=2$.}
\label{f:dyna_phase_trans}
\end{figure}

We present in  Fig.~\ref{f:dyna_phase_trans} the  comparison of these two linear instabilities   for the AP and AS models. The insets illustrate  the  typical static, stalled and oscillatory   regimes. Similar picture  emerges  from the study of the   linear stability of the stalled solutions \textcolor{black}{$(S_s^{\pm},C_s^{\pm})$}:
\begin{itemize}
\item If $k_{\text{p}}\leq k_{\text{p}}^*$, the configuration $(S_s^{\pm},C_s^{\pm})$ stops being linearly stable through a Hopf bifurcation taking place at $\P= \P_s- k_{\text{p}}(\P_s-\P_m)/k_{\text{p}}^*$.
\item If $k_{\text{p}}\geq k_{\text{p}}^*$, the configuration $(S_s^{\pm},C_s^{\pm})$ stops being linearly stable through a supercritical pitchfork bifurcation taking place at $\P=\P_m$. 
\end{itemize} 

\begin{figure}[h!]
\centering
\includegraphics[scale=0.6]{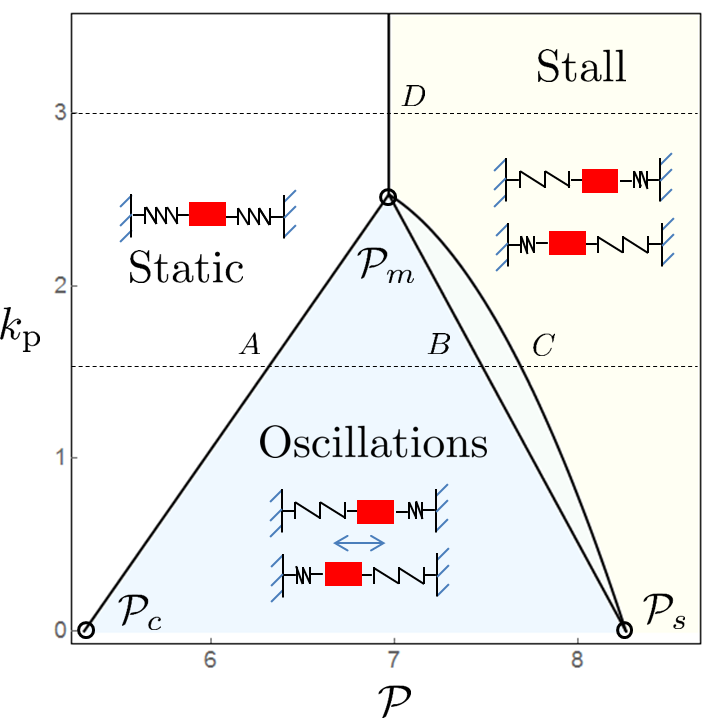}
\caption{\textcolor{black}{(Color online) Stability diagram of an AP confined by harmonic springs depending on the contractility and environment stiffness}. Note the region of coexistence (metastability) between the oscillations and stall phases where the two types of solutions coexist. The two thin dashed lines and associated capital letters are related to the bifurcations diagrams shown on Fig.\ref{f:bifdiag}. Non-dimensional length $\mathcal{L}=2$.}
\label{f:active_particle_stifness}
\end{figure}

Both \textcolor{black}{linear stability} results are summarized  for the AP model on the synthetic phase diagram shown in Fig.~\ref{f:active_particle_stifness}. Moreover, we show there the \textcolor{black}{numerically constructed non-linear} stability boundaries  for  all three   types of solutions: static, stalled, and oscillatory.   Interestingly, the supercritical transition from a static to a stalled state becomes insensitive to $k_{\text{p}}$ above the threshold $k_{\text{p}}^*$. This indicates again that the  AP  can self-adapt to the environmental stiffness in order to maintain  the same motor activity  threshold.  We also report the opening of a domain of metastability where oscillatory solutions coexist with stall solutions \textcolor{black}{which we were not able to capture numerically for the AS model in \cite{recho2019force}. This is potentially important as it can open the possibility of complex stop-and-go dynamics for an elastically confined AP subjected to noise.} 

\begin{figure}[h!]
\centering
\includegraphics[scale=0.5]{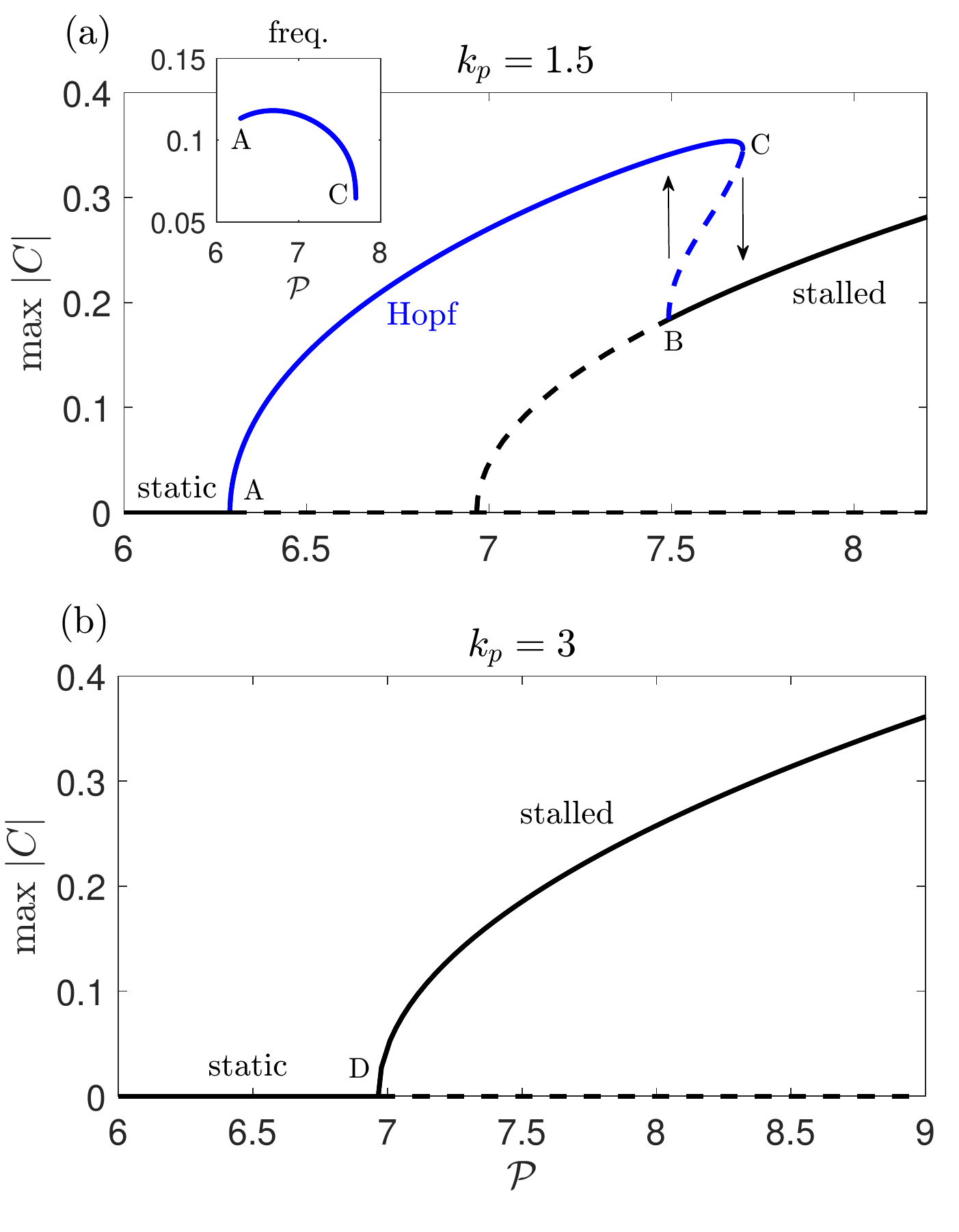}
\caption{ (Color online) Two typical bifurcation diagrams for an elastically confined AP \textcolor{black}{when the stiffness $k_{\text{p}}<k_{\text{p}}^*$ (a) and when $k_{\text{p}}>k_{\text{p}}^*$ (b).} The continuation of the Hopf bifurcation is shown in blue while supercritical pitchfork (second order phase transitions) are shown in black. In inset of (a), we show the frequency (inverse of the period) of the stable Hopf oscillations as a function of the continuation parameter. The dashed lines are linearly unstable branches while full lines are stable. Arrows indicate the discontinuous transitions. The capital letters are related to the phase diagram shown on Fig.~\ref{f:active_particle_stifness}. Non-dimesnional length $\mathcal{L}=2$. }
\label{f:bifdiag}
\end{figure}

The origin of such metastability can be understood by reconstructing the global structure of the bifurcation diagram using a numerical continuation method \cite{DoeKelKer_ijbc91}. Typical results are illustrated  in Fig.~\ref{f:bifdiag}. \textcolor{black}{When the environment stiffness is smaller than the tri-critical point value, i.e. $k_{\text{p}}\leq k_{\text{p}}^*$}, see  Fig.~\ref{f:bifdiag}~(a), the branch of oscillatory solutions  emerging from the static branch  reaches a turning point (denoted C in Fig.~\ref{f:active_particle_stifness} and Fig.~\ref{f:bifdiag}). As the \textcolor{black}{contractility} $\P$ increases beyond this point   the oscillatory solutions cease to be stable and the system abruptly switches  to the stalled configuration. The same discontinuous transition is associated with the  decrease of  \textcolor{black}{contractility} when the stalled configurations  undergoes  a Hopf bifurcation at  the critical value of  $\P$   (denoted by  B in Fig.~\ref{f:active_particle_stifness} and Fig.~\ref{f:bifdiag}) . The oscillatory  and stalled configurations have  therefore   a domain of metastable coexistence. When $k_{\text{p}}\geq k_{\text{p}}^*$, see Fig.~\ref{f:bifdiag}~(b), this complexity disappears as we only observe a continuous transition from a static to a stalled state  (Point D on Fig.~\ref{f:active_particle_stifness} and Fig.~\ref{f:bifdiag}).
 
\begin{figure}[h!]
\centering
\includegraphics[scale=0.5]{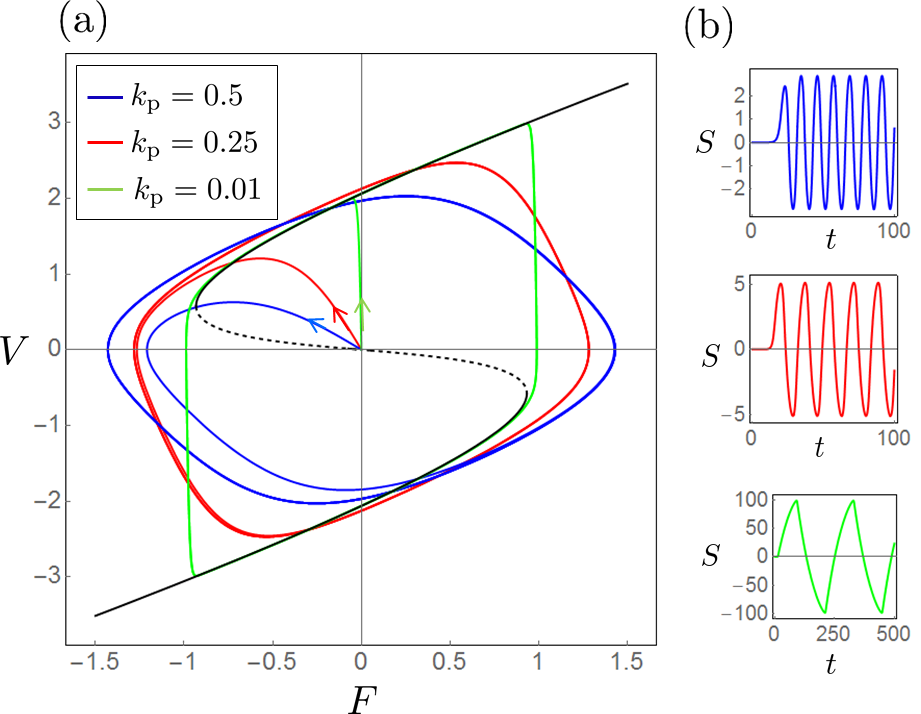}
\caption{(Color online) Dynamics of the oscillations when the stiffness  $k_{\text{p}}$ of the environment starting from the initial state $(S_0,C_0)$. (a) represents the trajectories of the oscillating particle in the phase space $(k_{\text{p}}S,\dot{S})$ for three different stiffnesses. We superimpose in black color the V-F curve obtained for a fixed force (Section.~\ref{sec:FV}).  On (b) we show the related particle position steady state oscillations. Non-dimensional length and contractility $\mathcal{L}=2$ and $\P=6.5$.}
\label{f:oscil_dyna}
\end{figure}

To illustrate   the structure of the oscillations we show in Fig. \ref{f:oscil_dyna}  the limit cycle type regimes for the AP model using the force-velocity coordinates.  The  parameters are chosen in the oscillatory phase shown in   Fig.~\ref{f:active_particle_stifness}. As the stiffness of the environment gets smaller, the oscillations amplitude and their period is increasing which results, in the limit $k_{\text{p}}\rightarrow 0$, in an almost steady state  behavior when  limit cycle is described by the hysteretic V-F relation obtained in Section.~\ref{sec:FV}. 

Oscillations driven by molecular motors are ubiquitous across various space and time scales in cell biology \cite{kruse2011spontaneous,blanchard2018pulsatile}. Cell shape oscillations are often shown to be resulting from a periodic regulation by signalling molecules (Rho GTPases) controlling the motors contractility \cite{nishikawa2017controlling}. Indeed, the activation/inhibition dynamic between several Rho GTPases can form an autonomous clock acting as a pacemaker \cite{qin2018biochemical}. However,  center of mass oscillations of living cells,  associated with periodic reversals of the molecular motors polarity,  were also repeatedly observed in experiment  \cite{lavi2016deterministic, godeau2016cyclic}.  There exist theoretical models of cell motility involving both cytoskeleton contraction and protrusion   aiming at capturing the emergence of such oscillations. For instance in  \cite{PhysRevLett.111.158102}, oscillations emerge from a coupling between cell shape and biochemical polarization while in \cite{2019arXiv190201730F}, they result from a delay between actin and myosin flow in the cell cortex. Here, we report that oscillations can also spontaneously arise from the mechanical interaction of the cell with its elastic environment. Interestingly, oscillations similar to the ones found in this paper,  were  also reported at a smaller scale where a bead-tailed actin filament propelled by the collective action of myosin motors was tethered to an optical trap \cite{PhysRevLett.103.158102}.

\begin{figure}[h!]
\centering
\includegraphics[scale=0.6]{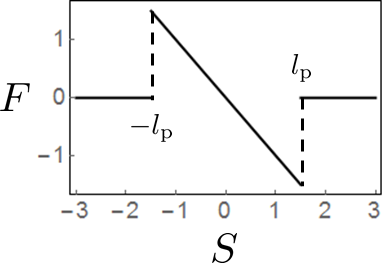}
\caption{Scheme of the harmonic confinement force with a threshold.}
\label{f:break_force_scheme}
\end{figure}

\begin{figure}[h!]
\centering
\includegraphics[scale=0.7]{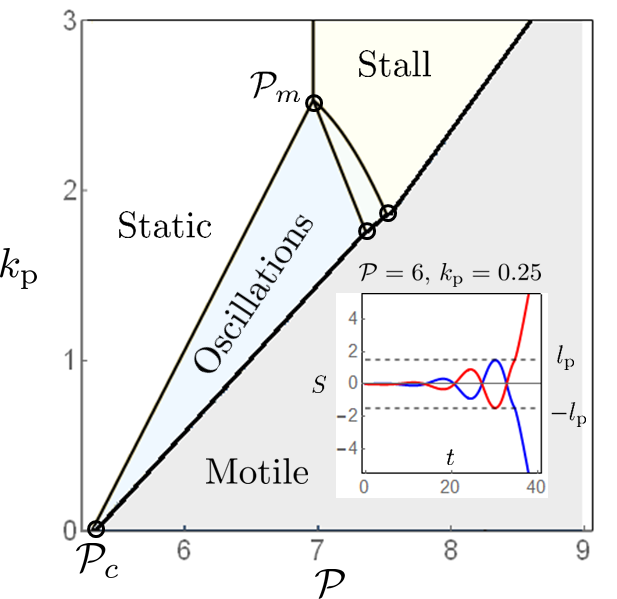}
\caption{(Color online) \textcolor{black}{Stability} diagram of an AP confined by breakable harmonic springs.  The position of the boundary of the motile phase depends on the initial conditions of  \eqref{eq:meta_model_231} which are here taken to be $(S,C)(t=0)=(S_0,C_0)$. We show in inset some typical dynamic of the AP breaking out of the harmonic confinement. Non-dimensional length and breaking limit $\mathcal{L}=2$ and $l_{\text{p}}=1.5$. }
\label{f:break_out}
\end{figure}

To complement this analysis, we now briefly discuss  the case of a breakable environmental confinement from where the active agent can escape. To this end we assume that   $ F(t)=-k_{\text{p}}S(t)H(l_{\text{p}}-|S(t)|),$ 
where the parameter $l_{\text{p}}$ characterizes the  (non-dimensional) breaking limit  of the confining spring, see Fig.~\ref{f:break_force_scheme}. 
In this case, the AP can break out of the confinement and reach a motile state. We show in Fig.~\ref{f:break_out} the resulting phase diagram at a given value of $l_{\text{p}}$.
The phase diagram in this case is shown in Fig.~\ref{f:active_particle_stifness}. We see that  the  AP can now become motile as soon as the spring reaches the elongation  $l_{\text{p}}$ with a non zero speed.   Such escape scenario is  reminiscent of an epithelial to mesenchimal transition where cells break out from the confinement of their neighbors and start to move persistently on their own.

\section{Conclusions}\label{sec:conclusion}

Starting with a one-dimensional model of contraction-driven  crawling we developed an  active particle model which is able to  adjust  its polarity to the applied force. Both models generate quantitatively similar force-velocity  relations which can describe  hysteresis in both velocity and force. In the presence of a viscous resistance from the environment, the obtained  model captures the emergence of polarity and the associated initiation of motility   when the viscosity is reduced.  If elastically confined,  both  active segment and particle  can develop dynamical oscillations. The model suggests that  there exists a domain of parameters where  oscillatory  and stalled states   coexist which suggests the possibility  of  stochastic switch between the two regimes.  We can anticipate even more  complex   dynamic attractors   in  a visco-elastic environment  of Kelvin-Voigt or Maxwell type and/or when the external rheology becomes non-linear involving, for instance, fracture or plastic deformations.   

\acknowledgments{ P.R. acknowledges support from a CNRS-Momentum grant. T.P. was supported by the EPSRC Engineering Nonlinearity project No. EP/K003836/1. L.T. is grateful to the French government which supported his work under Grant No. ANR-10-IDEX-0001-02 PSL.}

\appendix

 \section{Effective viscosity}\label{sec:appendix_B}
When the applied force is much bigger than the  contractile force  $\F\gg\mathcal{P}$, Eq.~\eqref{eq:velocity_fronts} furnishes the explicit steady-state velocity-force (V-F) relation:
$ V= (\F/2)\coth(\mathcal{L}/2).$
The inverse of the slope of the V-F relation at zero force (i.e. the effective frictional viscosity) can then be computed directly: 
$$\mu_{\infty}=2\tanh(\mathcal{L}/2).$$ 

In the opposite case, when the external forces are negligible $\mathcal{P}\gg F$, the homogeneous solution $c\equiv 1$ is the stable steady state as long as $\mathcal{P}\leq \mathcal{P}_c$. By performing a  first order Taylor expansion around this solution for small  $\F$,  we can compute the effective frictional viscosity $\mu_0$. 

To this end we introduce  a small parameter $\epsilon$ and substitute the expansions 
$c(y)=1+\epsilon c_1(y)\text{, }V=\epsilon V_1$ and  $\F=\epsilon \F_1$
into equation \eqref{eq:model}.  At the first order we obtain the linear integro-differential equation  
\begin{multline} \label{a1}
\int_{-1/2}^{1/2}\left[\phi(\mathcal{L}(u-v))-\phi(\mathcal{L}/2-\mathcal{L}v)\right] c_1(v)dv+\\\F_1[\psi(\mathcal{L}u)-\psi(\mathcal{L}/2)]=\frac{1}{\mathcal{L}}\partial_u c_1.
\end{multline}
Here we have used the rescaled variable $u$ and applied the no-flux boundary conditions.  
Note that we still need to impose the constraint $\int_{-1/2}^{1/2} c_1(u)du=0.$ 
\begin{figure}[!h]
\begin{center}
\includegraphics[scale=0.6]{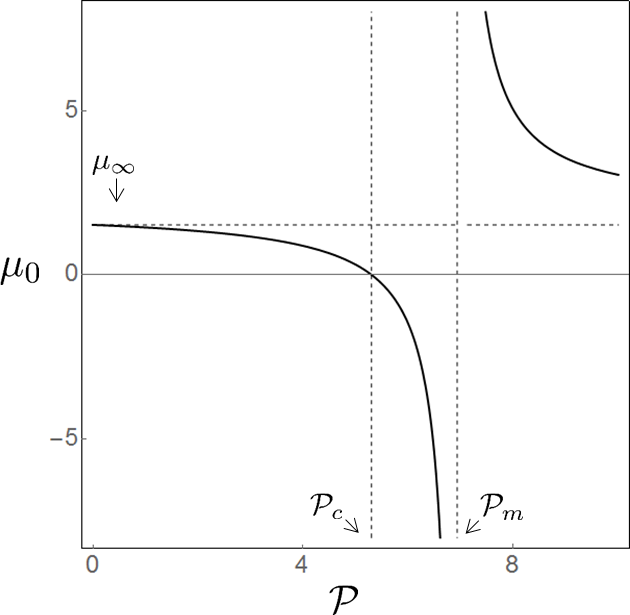}
\caption{\label{fig:slip} Effective frictional viscosity at the origin of the V-F curve, $\mu_{0}$ as a function of the non-dimensional contractility $\mathcal{P}$. Non-dimensional length $\mathcal{L}=2$.}
\end{center}
\end{figure} 

In view of the exponential nature of the kernel, the  equation \eqref{a1} can be solved analytically. 
It can be first transformed into the following system of second order linear differential equations
\begin{multline}
-\frac{1}{\mathcal{L}^2}\partial_{uu} \mathbf{X}+\mathbb{M}\mathbf{X}=\mathbf{V}(u),
\quad\text{where}\quad
\mathbf{X}=\left(\begin{array}{c}c_1\\\sigma_1\end{array}\right)\text{, }\\
\mathbf{V}=\left(\begin{array}{c}-\F_1 \tilde{\psi}(\mathcal{L}u)\\0\end{array}\right)\text{ and }
\mathbb{M}=\left(\begin{array}{cc}-\mathcal{P}/\mathcal{L}&1\\-\mathcal{P}/\mathcal{L}&1\end{array}\right).
\end{multline}
The boundary conditions take the form
\begin{multline}
\partial_u c_1\vert_{-1/2}=0\text{, }\int_{-1/2}^{1/2}c_1(u)du=0\text{, }\\
\sigma_1\vert_{-1/2}=\sigma_1\vert_{1/2}\text{ and } \partial_u\sigma_1\vert_{-1/2}=\partial_u\sigma_1\vert_{1/2}. 
\end{multline}
The solution of this system reads 
$$
c_1(u)=\frac{\csch\left(\omega/2\right) \sinh (u \omega )-u \omega  \coth\left(\omega/2\right)}{2-(\mathcal{P/L})  \omega  \coth \left(\omega/2\right)} F_1,
$$
where $\omega^2=\mathcal{L}^2\left(1- \mathcal{P}/\mathcal{L} \right)$. 
Finally, the substitution of $c_1(u)$ into \eqref{eq:velocity_fronts} gives the linear part of the force velocity relation 
\begin{equation}\label{eq:Force_velocity_small_force}
V_1=\F_1\left( \frac{\omega}{\mathcal{L}}\right)^3 \frac{\coth\left(\omega/2\right)}{2-(\mathcal{P/L})\omega\coth\left(\omega/2\right)}.
\end{equation}

 In Fig.~\ref{fig:slip} we show the effective viscosity at zero force 
$\mu_{0}= \partial \F_1/\partial V_1$ as a function of the parameter $\mathcal{P}$.
When $\mathcal{P}=0$, the value $\mu_{0}$ coincides with $\mu_{\infty}$ because the  V-F relation is linear over the whole range of forces. As the parameter $\mathcal{P}$ increases, $\mu_{0}$ decreases, which is the signature of the contractile activity being responsible for the induced flow inside the cell. When $\mathcal{P}$ reaches the value $\mathcal{P}_c$, the viscosity $\mu_{0}$ vanishes and becomes negative for $\mathcal{P}>\mathcal{P}_c$. The value of $\mu_{0}$ continues to decrease  and eventually diverges at  $\mathcal{P}=\mathcal{P}_m$ indicating a complete flattening of the V-F relation close to the origin. If $\mathcal{P}$   increases beyond $\mathcal{P}_m$, then $\mu_{0}$ becomes positive again. Note, however, that expression \eqref{eq:Force_velocity_small_force} is obtained as we perturbed the homogeneous solution.  It is the stable attractor of the initial value problem only when $\mathcal{P}\leq \mathcal{P}_c$, and therefore, the values of $\mu_{0}$ obtained for $\mathcal{P}\geq \mathcal{P}_c$  are associated with unstable regimes (when the external force is controlled.) 

\textcolor{black}{\section{Non-positive definiteness of the coefficient $k_S$}\label{sec:appendix_C}}

\textcolor{black}{ Using the known expression of the kernel $\phi$, the general  expression for the coefficient  $k_S$, given by \eqref{e:k_Sgeneric}, can be rewritten  in the form
\begin{equation} \label{33}
k_S=\int_{-\L/2}^{\L/2}\frac{\cosh(z)}{2\sinh(\L/2)}f(z)dz.
\end{equation}
It then clear that if $f(z)\geq 0$ then $k_S\geq 0$. However, if we use the  non-sign-definite distributed loading $f$  which  takes  negative values close to  the sides of the segments and positive value in the center, then   the structure of the explicit multiplier in front of $f(z)$ in \eqref{33}  suggests that $k_S$ can become negative. For  instance,   if $f(z)=-a[\delta(z+\L/2)+\delta(z-\L/2)]+b\delta(z)$ with $a,b\geq$ and $b-2a=1$, then   $k_S\leq0$  as long as  $a\geq 1/(4\sinh(\L/4)^2)$. }

\bigskip

\section{Bifurcations and normal forms}\label{sec:appendix_A}

Consider a steady-state configuration of AS in the absence of an externally applied force ($F=0$). In this case $\partial_tc\equiv 0$ and we can integrate \eqref{eq:motor_conserv} to obtain an expression for $c$ as a function of $\sigma$. Then substituting this expression into \eqref{eq:force_bal} leads to the non-local boundary value problem (see~\cite{RecPutTru_jmps15} for details):
\begin{equation}\label{systTW}
\left\lbrace \begin{array}{c}
-\frac{1}{\mathcal{L}^2}\partial_{uu}s(u)+s(u)+s^b=\theta\left( \frac{e^{s(u)-\nu u}}{\int_{-1/2}^{1/2}e^{s(u)- \nu u}du}-1\right) \\
s(\pm\frac{1}{2})=0 \text{ and } \partial_us(\pm\frac{1}{2})=\nu,
\end{array}\right.
\end{equation}
Here we introduced the notations  $u=y/\mathcal{L}\in [-1/2,1/2]$,
$\theta\equiv\mathcal{P}/\mathcal{L}$, $s(u)=\sigma(u)-\sigma^b$, $s^b=\sigma^b-\theta$ and $\nu=\mathcal{L}V$. 
For steady states and $F=0$, equation \eqref{systTW} is equivalent to \eqref{eq:model} as $c$ can be reconstructed from $s$ and $\nu$ using the formula
$$
c(u)=\frac{e^{s(u)-\nu u}}{\int_{-1/2}^{1/2}e^{s(u)- \nu u}du}.
$$
Eq.~\eqref{systTW} has the unique homogeneous solution 
$$
s=0\text{, } \nu=0\text{ and } s^b=0.
$$
Below we study the  bifurcations from this  trivial solution giving rise to  nontrivial solutions as the parameter $\theta$ increases.

To this end, we choose a small parameter $\epsilon$  and  perform a Taylor expansion near the homogeneous solution keeping the terms up to third order: 
$$
\begin{cases}
s=0+\epsilon s_1+\frac{\epsilon^2}{2}s_2+\frac{\epsilon^3}{6}s_3+o(\epsilon^3),\\
\nu=0+\epsilon\nu_1+\frac{\epsilon^2}{2}\nu_2+\frac{\epsilon^3}{6}\nu_3+o(\epsilon^3),\\
s^b=0+\epsilon {s^b_1}+\frac{\epsilon^2}{2}{s^b_2}+\frac{\epsilon^3}{6}{s^b_3}+o(\epsilon^3).
\end{cases}
$$
We also write a similar expansion for  the bifurcation parameter  
$$\theta=\theta_0+\epsilon\theta_1+\frac{\epsilon^2}{2}\theta_2+\frac{\epsilon^3}{6}\theta_3+o(\epsilon^3).$$
Substituting these expansions into \eqref{systTW} and introducing the operator,
$$\mathbb{L}_{\text{in}}(s(u),s^b,\nu)=-\frac{\partial_{uu}s(u)}{\mathcal{L}^2}+(1-\theta_0)s(u)+(1-\theta_0)s^b+\theta_0 \nu u,$$
we obtain:
 
\noindent$\bullet$ at first order
\begin{equation}\label{firstorder}
\mathbb{L}_{\text{in}}(s_1,{s^b_1},\nu_1)=0,
\end{equation}
$\bullet$ at second order,
\onecolumngrid
\begin{multline}\label{secondorder}
\mathbb{L}_{\text{in}}(s_2,{s^b_2},\nu_2)=2 \left(\frac{1}{24} \theta_0 \left(12 s_1(u) (s_1(u)+2 {s^b_1}-2 \nu_1 u)+24 \nu_1\int_{-1/2}^{1/2} u s_1(u) \, du\right.\right.\\
\left. \left. -12 \int_{-1/2}^{1/2} s_1(u)^2 \, du+24 {s^b_1}^2-24 {s^b_1} \nu_1 u+12 \nu_1^2 u^2-\nu_1^2\right)+\theta_1
   (s_1(u)+{s^b_1}-\nu_1 u)\right),
\end{multline}
\twocolumngrid
$\bullet$ at third order,
\onecolumngrid
\begin{multline}\label{thirdorder}
\mathbb{L}_{\text{in}}(s_3,{s^b_3},\nu_3)=\frac{1}{4} \left(\theta_0 \left(12 s_1(u) \left(2 {s^b_1}^2+{s^b_2}\right)+12 s_2(u)(s_1(u)+{s^b_1}-\nu_1 u)\right.\right.\\
\left. \left.+12\left(\int_{-1/2}^{1/2} u s_1(u) \, du\right) \left(2 \nu_1 s_1(u)+4 {s^b_1} \nu_1-2 \nu_1^2 u+\nu_2\right)-24 {s^b_1} \nu_1u s_1(u)+12 {s^b_1} s_1(u)^2\right.\right.\\
\left. \left.-24 {s^b_1} \left(\int_{-1/2}^{1/2} s_1(u)^2 \, du\right)-12 \int_{-1/2}^{1/2} s_1(u) s_2(u) \, du+12 \nu_1^2 u^2s_1(u)-12 \nu_1^2 \int_{-1/2}^{1/2} u^2 s_1(u) \, du-\nu_1^2 s_1(u)\right.\right.\\
\left. \left.-12 \nu_1 u s_1(u)^2+12\nu_1 u \int_{-1/2}^{1/2} s_1(u)^2 \, du+12\nu_1 \int_{-1/2}^{1/2} u s_1(u)^2 \, du-12\nu_2 u s_1(u)+4 s_1(u)^3-4 \int_{-1/2}^{1/2} s_1(u)^3 \, du\right.\right.\\
\left. \left.-12 s_1(u)\left(\int_{-1/2}^{1/2} s_1(u)^2 \, du\right)+24 {s^b_1}^3+u \left(-24 {s^b_1}^2 \nu_1-12 {s^b_1} \nu_2-12 {s^b_2} \nu_1+\nu_1^3\right)+24 {s^b_1} {s^b_2}-2{s^b_1} \nu_1^2\right.\right.\\
\left. \left.+12 \nu_1 u^2 ({s^b_1} \nu_1+\nu_2)+12 \nu_1 \int_{-1/2}^{1/2} u s_2(u) \, du-4 \nu_1^3 u^3-\nu_1 \nu_2\right)\right.\\
\left.+\theta_1 \left(12 s_1(u)(s_1(u)+2 {s^b_1}-2 \nu_1 u)+24 \nu_1 \int_{-1/2}^{1/2} u s_1(u) \, du-12 \int_{-1/2}^{1/2} s_1(u)^2 \, du+24 {s^b_1}^2-12 u (2 {s^b_1}\nu_1+\nu_2)\right.\right.\\
\left.\left.+12 s_2(u)+12 {s^b_2}+12 \nu_1^2 u^2-\nu_1^2\right)+12 \theta_2 (s_1(u)+{s^b_1}-\nu_1 u)\right).
\end{multline}
\twocolumngrid
At all orders the boundary conditions remain
\begin{equation}\label{genericbc}
s_i(\pm1/2)=0\text{ and } \partial_us_i(\pm1/2)=\nu_i.
\end{equation}

\subsection{Bifurcation points}\label{sec:linana}
The spectral analysis of \eqref{firstorder} produces a countable number of bifurcation points. Although we provide below a general analysis of all these points, we emphasis that direct numerical simulations of \eqref{eq:model} show that the only stable steady state branches are the trivial branch when $\mathcal{P}\leq \mathcal{P}_c$ and the first motile branch $D_1$ when  $\mathcal{P}> \mathcal{P}_c$. See \cite{RecPutTru_jmps15} for further details.

Introducing 
\begin{equation}\label{e:def_omega}
\omega^2=\mathcal{L}^2(1-\theta_0),
\end{equation}
we obtain in the first order,
\begin{multline}
s_1(u)=C_1 \cosh \left[\omega \left(u+1/2\right)\right]+C_2 \sinh \left[\omega \left(u+1/2\right)\right]\\
-s^b_1 + \nu_1 u \left(\omega^2-\mathcal{L}^2\right)/\omega^2 .
\end{multline}
Note that the solution $\omega=0$ should be excluded because it produces the same homogeneous solution. 
The four constants $C_1$, $C_2$, ${s^b_1}$ and $\nu_1$ follows from the four boundary conditions \eqref{genericbc}, 
which leads to a homogeneous linear system of equations. This algebraic problem has nontrivial solutions when 
the determinant of the matrix
\begin{equation}\label{matristiff}
\left(
\begin{array}{cccc}
 1 & 0 & -1 & \left(\mathcal{L}^2/\omega^2-1\right)/2 \\
 0 & \omega & 0 & -\mathcal{L}^2/\omega^2 \\
 \cosh (\omega) & \sinh (\omega) & -1 & \left(1-\mathcal{L}^2/\omega^2\right)/2 \\
 \omega \sinh (\omega) & \omega \cosh (\omega) & 0 & -\mathcal{L}^2/\omega^2 \\
\end{array}
\right)
\end{equation}
cancels out, yielding the transcendental characteristic equation
$$
2[\cosh(\omega)-1]+(\omega^2/\mathcal{L}^2-1) \omega \sinh(\omega)=0.
$$
The solutions of this equation split into two families depending on whether parameter $\omega$ is real or purely imaginary. 
In the first (resp. second) case we denote $\omega_c=|\omega|\geq 0$ (resp. $\omega_c=-|\omega|\leq 0$), which leads to  
\begin{equation}\label{bifurc_val_bis}
\left\{ \begin{array}{c}
2\tanh(\omega_c/2)=(1-\omega_c^2/\mathcal{L}^2)\omega_c \text{ if $\omega_c\geq 0$ } \\
2[\cos(\omega_c)-1]+(\omega_c^2/\mathcal{L}^2+1)\omega_c\sin(\omega_c)=0 \text{ if $\omega_c\leq 0$ } 
\end{array} \right.
\end{equation}
It is convenient to analyze equations \eqref{bifurc_val_bis}$_1$ and \eqref{bifurc_val_bis}$_2$ separately:
\begin{enumerate}
\item When $\omega$ is real, equation \eqref{bifurc_val_bis}$_1$ has a unique solution provided
$2\sqrt{3}\leq \mathcal{L}$.
Otherwise, it has no solution. The corresponding  eigenvector can be written as
$$
\begin{array}{c}
\left(
\begin{array}{c}
 {s^b_1}\\
  \nu_1\\
  s_1(u)
\end{array}
\right)
= 
\left(
\begin{array}{c}
0\\
1\\
\frac{\mathcal{L}^2}{\omega_c^3\cosh(\omega_c/2)}\left[ \sinh\left(u \omega_c\right)-2u \sinh \left(\omega_c/2\right)\right]
\end{array}
\right)
\end{array}
$$
Since $\nu_1\neq 0$ the corresponding bifurcation leads to a motile configuration  that we denote $D_1$.  

\item When $\omega$ is purely imaginary, equation \eqref{bifurc_val_bis}$_2$ has two families of solutions:
\begin{enumerate}
\item The first family is explicitly parametrized with an integer $\omega_c=-2m\pi$ with $m\geq 1$ and
the associated eigenvector reads
$$
\left(
\begin{array}{c}
  {s^b_1}\\
  \nu_1\\
  s_1(u)
\end{array}
\right)
= \left(
\begin{array}{c}
1\\
0\\
\cos[\omega_c (u+1/2)]-1
\end{array}
\right)
$$
Since $\nu_1=0$, the bifurcated solution describes a static cell. 
We denote this family $S_m$. 

\item The second family consists of a countable set of negative roots of the equation
\begin{equation}\label{implicitomcbis}
2\tan(\omega_c/2)=(1+\omega_c^2/\mathcal{L}^2)\omega_c
\end{equation}
The largest root exists if only if $\mathcal{L} \leq 2\sqrt{3}$ and the corresponding eigenvector reads
$$
\begin{array}{c}
\left(
\begin{array}{c}
  {s^b_1}\\
  \nu_1\\
  s_1(u)
\end{array}
\right)= 
\left(
\begin{array}{c}
0\\
1\\
\frac{-\mathcal{L}^2}{\omega_c^3\cos(\omega_c/2)}\left[ \sin\left(u\omega_c\right)-2u\sin \left(\omega_c/2\right)\right]
\end{array}
\right)
\end{array}
$$
Since $\nu_1\neq 0$, these roots of the characteristic equation are associated with motile branches. We denote this family $D_m$  with $m\geq 1$.
\end{enumerate}
\end{enumerate}

The critical bifurcation threshold $\mathcal{P}_c$ introduced in the main text can be written as $\mathcal{P}_c=\mathcal{L}\theta_0(D_1)$ where the relation between $\theta_0$ and $\omega_c$ follows from \eqref{e:def_omega}. See also Fig.~\ref{D1shape}. 

In the presence of an external viscous friction $\eta_\text{p}$, the Dirichlet boundary conditions in \eqref{systTW} are modified into 
$$s(\pm\frac{1}{2})=\mp\frac{\eta_\text{p}\nu}{2\L},$$
which modifies the critical contractility value controlling the onset of motility  into $\mathcal{P}_c^{\text{eff}}(\L, \eta_{\text{p}})$ as \eqref{bifurc_val_bis}$_1$ and \eqref{implicitomcbis} respectively become
$$2\tanh(\omega_c/2)=[1-\omega_c^2/\mathcal{L}^2(1+\eta_{\text{p}}/\L)]\omega_c $$
\centerline{and}
$$2\tan(\omega_c/2)=[1+\omega_c^2/\mathcal{L}^2(1+\eta_{\text{p}}/\L)]\omega_c.$$
The resulting value of $\mathcal{P}_c^{\text{eff}}$ is shown on Fig.~\ref{f:active_particle_viscosity}.

\subsection{Normal forms}
Each bifurcation is now characterized by the eigenvalue $\theta_0$  and the eigenvector $[s_1(u),{s^b_1},\nu_1]$. 
This information is not sufficient to find the shape of the bifurcated branch close to a bifurcation point. 
To this end we need to use higher order equations \eqref{secondorder}-\eqref{thirdorder}.

Starting from second order, the right-hand side of equation \eqref{secondorder} must be in the range of the operator~$\mathbb{L}_{\text{in}}$. This  is equivalent (Fredholm alternative) to the requirement that this expression is orthogonal to the kernel of the dual of~$\mathbb{L}_{\text{in}}$. In our case, this property reduces to imposing an orthogonality condition in the space $(C_1,C_2,s^b,\nu)$ with the kernel of the transpose of \eqref{matristiff}. The resulting scalar equation sets the value of $\theta_1$. 

For both static and motile branches, we find $\theta_1=0$, which means that the static and motile bifurcations are of pitchfork type. 
The super- or sub-critical nature of the bifurcation follows from third order. Solving first \eqref{secondorder} with $\theta_1=0$ leads to 
the solution $[s^b_2,\nu_2,s_2(u)]$, whose detailed expression is not given here. 

\begin{figure}[h!]
\includegraphics[scale=0.45]{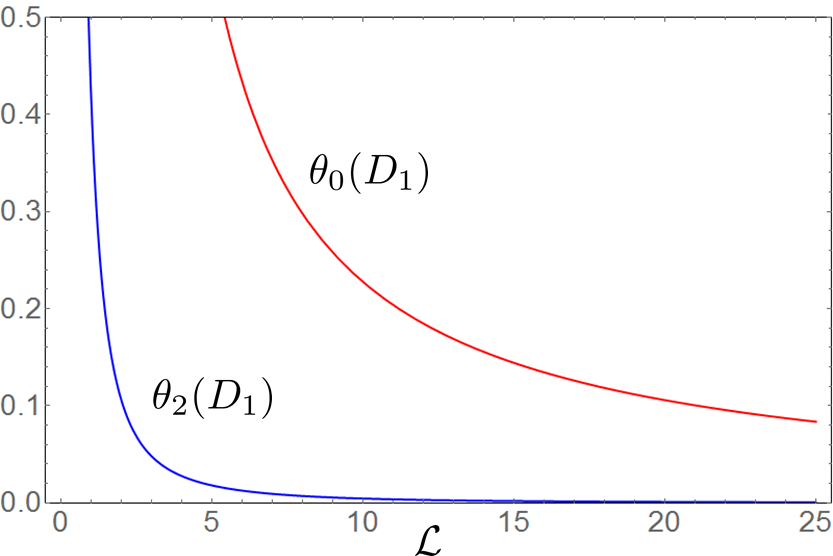}
\caption{\label{D1shape} Eigenvalues $\theta_0$ and $\theta_2$ as a function of $\mathcal{L}$ for the first motile branch $D_1$.}
\end{figure}
We can now apply the same analysis as above to equation \eqref{thirdorder} which gives, for the motile branches,
\onecolumngrid 
\begin{multline}
\theta_2= \left[\left(\mathcal{L}^2-\omega^2\right) \left(\mathcal{L}^{10} \left(3 \omega^2+770\right)-6 \mathcal{L}^8 \left(2 \omega^4+215 \omega^2+1540\right)+6 \mathcal{L}^6 \left(4 \omega^4+85 \omega^2+660\right) \omega^2\right. \right. \\
\left. \left. -2 \mathcal{L}^4 \left(15 \omega^2+79\right) \omega^6+21 \mathcal{L}^2 \left(\omega^2+8\right) \omega^8-6
   \omega^{12}\right)\right]/\left[ 144 \mathcal{L}^2 \omega^8 \left(-\mathcal{L}^4+2 \mathcal{L}^2 \left(\omega^2+6\right)-\omega^4\right)\right] 
\end{multline}
\twocolumngrid
This expression is always positive indicating that all motile bifurcations are supercritical. 
We illustrate in Fig.~\ref{D1shape} the eigenvalues $\theta_0$ and $\theta_2$ as a function of the length parameter~$\mathcal{L}$ 
for the first motile branch $D_1$. 

\begin{figure}[!h]
\begin{center}
\includegraphics[scale=0.45]{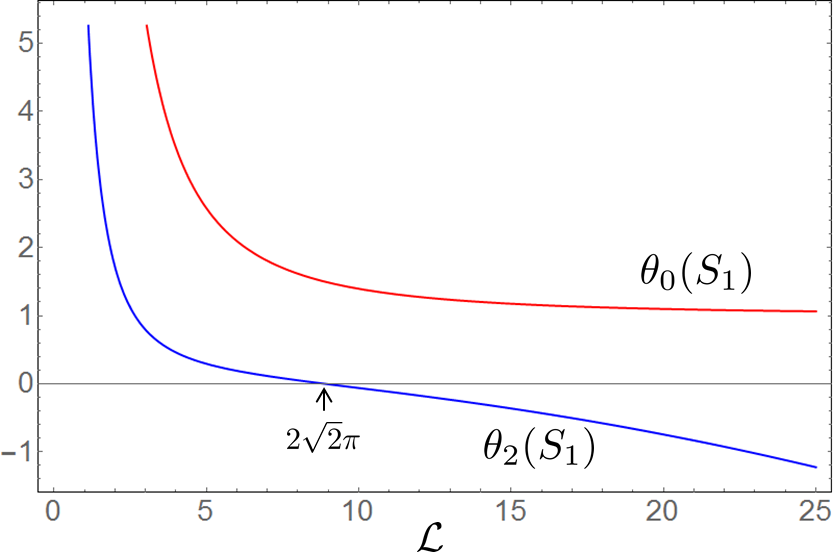}
\caption{\label{S1shape} Eigenvalues $\theta_0$ and $\theta_2$ as a function of $\mathcal{L}$ for the first static branch $S_1$. }
\end{center}
\end{figure}

To complete the picture, a similar but simpler analysis for the static branches can be carried out. 
Given that the expression for $\theta_0$ can be given explicitly, we can compute 
$$
\theta_2=\frac{1}{48} \left(\frac{32 \pi ^2 m^2}{\mathcal{L}^2}-\frac{\mathcal{L}^2}{\pi ^2 m^2}+4\right) .
$$
This value is not always positive which indicates that the pitchfork bifurcation can be super- or sub-critical depending on the value of $m$ and $\mathcal{L}$. In Fig.~\ref{S1shape}, we illustrate the dependence of $\theta_0$ and $\theta_2$ on $\mathcal{L}$ for the first static branch $S_1$ ($m=1$).

\bibliographystyle{apsrev4-1}

\clearpage

\end{document}